\documentclass[acmlarge,screen,authordraft,nonacm,review=false,timestamp=false]{acmart}

\setcopyright{acmlicensed}
\copyrightyear{2026}
\acmYear{2026}
\acmDOI{XXXXXXX.XXXXXXX}
\acmJournal{IMWUT}
\acmVolume{X}
\acmNumber{X}
\acmArticle{XXX}
\acmMonth{X}

\usepackage{xcolor}
\usepackage{subcaption}
\usepackage{tabularray}
\UseTblrLibrary{booktabs}
\usepackage{cleveref}

  \newcommand{\system}{UWB-Fat} 

\begin{document}

\title{\system: Non-Intrusive Body Fat Measurement Using Commodity Ultra-Wideband Radar}

\author{Haotang Li}
\affiliation{
  \institution{University of Arizona}
  \city{Tucson}
  \country{United States}
}
\email{haotangl@arizona.edu}

\author{Yili Ren}
\affiliation{
  \institution{University of South Florida}
  \city{Tampa}
  \country{United States}
}
\email{yiliren@usf.edu}

\author{Zhenyu Qi}
\affiliation{
  \institution{University of Arizona}
  \city{Tucson}
  \country{United States}
}
\email{qzydustin@arizona.edu}

\author{Sen He}
\affiliation{
  \institution{University of Arizona}
  \city{Tucson}
  \country{United States}
}
\email{senhe@arizona.edu}

\author{Kebin Peng}
\affiliation{
  \institution{East Carolina University}
  \city{Greenville, North Carolina}
  \country{United States}
}
\email{pengk24@ecu.edu}

\author{Sheng Tan}
\affiliation{
  \institution{Trinity University}
  \city{San Antonio}
  \country{United States}
}
\email{stan@trinity.edu}

\author{Bo Liu}
\affiliation{
  \institution{University of Arizona}
  \city{Tucson}
  \country{United States}
}
\email{boliu@arizona.edu}

\author{Jiyue Zhao}
\affiliation{
  \institution{The University of Georgia}
  \city{Athens, Georgia}
  \country{United States}
}
\email{jzhao@uga.edu}

\author{Zi Wang}
\affiliation{
  \institution{Augusta University}
  \city{Augusta}
  \country{United States}
}
\email{zwang1@augusta.edu}

\begin{abstract}

Body fat percentage and its spatial distribution are clinically important health indicators. However, existing measurement methods often impose a tradeoff between accuracy and accessibility. Clinical-grade techniques, such as Dual-Energy X-ray Absorptiometry (DEXA) and hydrostatic weighing, provide accurate measurements but require specialized equipment and trained operators, making them difficult to access and unsuitable for everyday use.
In contrast, consumer-level methods, such as Bioelectrical Impedance Analysis (BIA) smart scales and skinfold calipers, are more accessible but typically provide only coarse-grained estimates, are prone to user error, or require intrusive physical contact.
In this work, we present \system{}, the first system that leverages commodity ultra-wideband (UWB) radar to enable non-intrusive, accessible, and accurate caliper-equivalent skinfold thickness estimation, serving as a convenient replacement for the skinfold caliper. \system{} collects UWB signal at specified body sites non-intrusively without operator assistance. It extracts body-composition-related features from UWB signals by exploiting dielectric contrasts among skin, fat, and muscle tissues. Then, it uses a physics-inspired model to estimate site-specific skinfold thickness. We evaluate \system{} on 15 participants, achieving a root mean square error of 0.63~mm for pooled-site subcutaneous fat thickness. These results highlight the potential of \system{} to support low-cost, self-administered, and everyday body fat monitoring.

\end{abstract}



\maketitle

\section{Introduction}
\label{sec:intro}

Body fat is one of the strongest modifiable predictors of metabolic disease, cardiovascular events, and all-cause mortality~\cite{jayedi2022bodyfat}. Accurate assessment of body fat is therefore important for both individual health management and population-level prevention. Yet the metric most widely used in clinical and consumer practice, body mass index (BMI)~\cite{BMI}, is only an indirect proxy for adiposity: it conflates fat mass with lean mass and cannot quantify either the amount or the anatomical distribution of adipose tissue~\cite{prentice2001beyond}. A more direct metric is total body fat percentage~\cite{BodyFat}, which estimates the proportion of body mass composed of fat. However, even this metric remains clinically incomplete when reported as a single whole-body value, because it does not reveal where fat is stored. This distinction is important because fat accumulated in different anatomical regions is associated with substantially different metabolic and cardiovascular risk profiles~\cite{RN19}.

This limitation motivates attention to regional fat distribution, which provides clinically important information beyond BMI and total body fat percentage. For example, abdominal adiposity is clinically important in part because it is associated with visceral adipose tissue, which accumulates around internal organs and is more strongly associated with type 2 diabetes and prediabetes than BMI~\cite{jung2016visceral}. Index-based evidence further supports this association: each unit increase in the visceral adiposity index has been independently associated with a 43\% increase in the odds of developing type 2 diabetes~\cite{zhou2024visceral}. Subcutaneous truncal fat is associated with elevated cardiovascular disease risk, whereas preferential fat deposition in the gluteofemoral region is associated with a comparatively lower metabolic risk profile~\cite{tchernof2013pathophysiology, manolopoulos2010gluteofemoral}. Thus, individuals with identical total body fat percentages may still face substantially different disease risks depending on their regional fat distribution. These findings underscore the need for accessible methods that can estimate body fat across different anatomical sites, rather than reporting only a single whole-body value.

Existing body fat measurement approaches available to users today require tradeoffs among accuracy, cost, accessibility, anatomical specificity, and measurement intrusiveness. No existing approach simultaneously provides accurate, low-cost, accessible, site-specific, and non-intrusive body fat measurement. Specifically,
\emph{Clinical gold-standard methods}, including Dual-Energy X-ray Absorptiometry (DEXA), hydrostatic weighing, and air-displacement plethysmography ~\cite{demerath2002comparison, levenhagen1999comparison, zanini2015body}, provide accurate body composition measurements. DEXA can further provide site-specific fat estimates, whereas hydrostatic weighing and air-displacement plethysmography primarily provide whole-body estimates. However, these methods require costly equipment and trained technicians, making them difficult to access and unsuitable for everyday use. In addition, DEXA exposes users to ionizing radiation, while hydrostatic weighing requires full-body submersion~\cite{messina2020dxa, fields2002airdisplacement}.
\emph{Magnetic resonance imaging (MRI)~\cite{thomas1998magnetic, muller2011quantification, baum2016mr} and computed tomography (CT)}~\cite{grauer1984quantification, kim2013body, sjostrom1986determination} provide reference-grade imaging for quantifying regional fat. However, they require specialized imaging facilities and are substantially more expensive than consumer-level approaches, limiting their suitability for routine body-composition monitoring. CT also exposes users to a relatively high dose of ionizing radiation~\cite{shen2003adipose}.
\emph{Consumer Bioelectrical Impedance Analysis (BIA) smart scales} are inexpensive and easy to use at home. However, they typically report only a coarse-grained whole-body fat percentage, are sensitive to factors such as hydration status and meal timing~\cite{kyle2004bioelectrical}.
\emph{Skinfold calipers} provide low-cost, site-specific estimates of subcutaneous fat thickness~\cite{machado2025skinfold}. However, this accessibility comes at the cost of intrusiveness and operator dependence: the procedure requires pinching a fold of skin with mechanical jaws, and accurate measurement typically requires a trained second person. In addition, skinfold measurements are prone to user error due to incorrect anatomical site selection, inconsistent measurement technique, and improper calibration of caliper pressure.

Ultra-wideband (UWB) radar~\cite{taylor2001ultra, barrett2001history, cheraghinia2024comprehensive, taylor2012introduction} offers a compelling alternative for non-intrusive body fat measurement. By transmitting ultra-short pulses and analyzing the reflected waveforms, UWB can capture fine-grained range profiles that are sensitive to tissue interfaces within the human body. Because skin, fat, muscle, and other tissues exhibit different dielectric properties at GHz frequencies, UWB reflections can encode information related to layered body composition~\cite{gabriel1996dielectric}. 
Importantly, UWB is no longer confined to specialized laboratory platforms. UWB radios have been increasingly integrated into consumer devices, including smartphones~\cite{heinrich2023smartphones,devrio2023smartposer}, smartwatches~\cite{cao2024uwb}, and item trackers~\cite{hany2024airtags}, for applications such as ranging, localization, and device-to-device interaction~\cite{malajner2015uwb, zwirello2012uwb, wymeersch2012machine, minoli2018ultrawideband}. Although current consumer UWB hardware is not designed for biomedical sensing, this rapid proliferation demonstrates a clear trend toward low-cost, compact, and widely deployable UWB platforms. Motivated by this trend, we investigate whether commodity-grade UWB radar can enable accessible, non-intrusive, and site-specific body fat estimation.

We present \system, the first system to estimate site-specific caliper-equivalent skinfold thickness from commodity UWB radar, serving as a non-intrusive alternative to skinfold calipers.
\system{} introduces three capabilities unavailable in prior approaches: (1) \emph{anatomical site-specific skinfold thickness measurement}, enabling fat estimation at relevant anatomical sites; (2) \emph{non-intrusive measurement}, requiring users only to place the UWB device on the target body site, without skin pinching or clamping; and (3) \emph{single-user, everyday operation}, allowing users to perform measurements at home without specialized equipment or assistance from a second operator.

Translating commodity UWB radar into accurate fat thickness measurement involves three technical challenges, and \system{} addresses each with a corresponding design choice. 
First, the bandwidth of consumer-grade UWB radios yields centimeter-scale time-domain resolution in fat, well above the 5 to 30\,mm physiological range, so peak-finding on then Channel Impulse Response (CIR) magnitude cannot separate the skin-fat and fat-muscle echoes. 
\system{} resolves this by reading fat thickness from the in-band complex baseband phase, where in the layered three-medium reflection model fat thickness enters the channel response only through a single phase propagator whose in-band slope is monotonic in thickness and observable without phase unwrapping. 
Second, the same in-band response carries nuisance parameters such as skin thickness and per-subject dielectric dispersion, all of which confound a closed-form readout of any single derived feature. 
\system{} adopts Full Waveform Inversion (FWI)~\cite{tarantola1984inversion, pratt1999seismic, rubaek2007nonlinear, mojabi2009overview} and fits the parameters of the physics-inspired model jointly to the broadband measurement, so that the nuisance parameters absorb their share of the residual rather than bleeding into the fat estimate. 
Third, anatomical and deployment variation, including different skin thicknesses across body sites, body-curvature differences, and imperfect normal placement of the radar, can change the measured response without changing the underlying fat thickness. 
\system{} mitigates these factors with site-aware anatomical priors that constrain the FWI search space, short-window temporal averaging that reduces frame-level noise, and a broadband data-misfit objective that remains robust to small antenna tilt. 
We instantiate these design choices as a deterministic signal-processing pipeline on top of a commodity Novelda X7F202 UWB radar module that transmits at a 7.875\,GHz carrier with a 460\,MHz processing bandwidth. 
Raw complex baseband CIR frames from the radar are coherently averaged at each anthropometric site, transformed into the in-band channel response \(\Gamma(f)\), and passed to the physics-inspired model. 
The solver jointly recovers skin thickness, fat thickness, and the antenna-skin air gap by minimizing the broadband data misfit against the closed-form layered-tissue forward model derived in \Cref{sec:prelim:tissue}. 
The recovered fat thickness is the system output, and the same configuration applies across body sites by selecting site-specific anatomical priors.

We evaluate this pipeline in a user study with 15 participants spanning a range of body types. 
At each session, we collect paired UWB radar measurements and skinfold-caliper ground truth at standard anthropometric sites including the abdomen, suprailiac, triceps, and anterior thigh. 
To characterize robustness, we additionally vary seven deployment factors across a subset of measurements, capturing realistic placement and posture variation rather than only ideal lab conditions.

This paper makes the following contributions:
\begin{itemize}
    \item We demonstrate, the first commodity-UWB system that recovers in-vivo, site-specific, caliper-equivalent skinfold thickness at accuracy comparable to caliper measurements. \system{} attains a pooled mean absolute error of $0.54$~mm in skinfold thickness across $15$ participants and $5$ anatomical sites.
  \item We introduce a body-site-aware fat sensing methodology that maps UWB reflections to standard anthropometric sites, enabling non-intrusive subcutaneous fat estimation at locations directly comparable to skinfold caliper measurements.
  \item We show that \system{} operates without skin pinching, without current injection, and without a second operator, and can be embedded in ambient scenarios using off-the-shelf UWB hardware already present in commercial devices.
  \item We release a labeled dataset of UWB radar recordings paired with skinfold caliper ground-truth measurements across 15 participants to support reproducibility and future research.
\end{itemize}

\section{Related Work} \label{sec:related}

\subsection{Body Composition Measurement Methods} \label{sec:related:methods}

DEXA measures bone mineral density, lean mass, and fat mass from differential X-ray attenuation at two energies, providing whole-body and regional estimates with high precision \cite{messina2020dxa}. Hydrostatic weighing and air displacement plethysmography (BodPod) derive fat mass from body density via Archimedes' principle \cite{fields2002airdisplacement}. All three require expensive equipment, trained operators, and scheduled clinic visits, precluding routine or continuous use.

BIA devices infer body composition from the electrical impedance of body tissues, which varies with fat content due to fat's low water and electrolyte content \cite{kyle2004bioelectrical}. While inexpensive and easy to use at home, consumer BIA scales return only a coarse whole-body fat percentage, and accuracy is sensitive to hydration status, meal timing, skin temperature, and electrode placement \cite{dehghan2008bia}. Segmental BIA devices, which measure impedance separately at the trunk, arms, and legs, improve regional specificity but still pass electrical current through the body and require firm skin-electrode contact and a stylized stance. \system{} complements consumer BIA by resolving fat at specific anatomical sites without injecting current or requiring a stylized measurement posture, and without calibration prerequisites tied to hydration. BIA addresses a different question than \system{}: BIA reports whole-body fat fraction, whereas \system{} recovers site-specific subcutaneous fat thickness. The two methods are therefore complementary rather than competing. 
Streeter et al.~\cite{streeter2024classification} studied microwave interrogation of multi-layer tissue-mimicking dielectric stacks from 2 to 20 GHz. Their work shows that broadband complex reflection measurements contain information about layered dielectric structure, and models the stack using transmission-line representations. \system{} builds on the same physical intuition that layered tissue interfaces are observable in broadband RF measurements, but addresses a different sensing problem. Streeter et al. classify a finite set of phantom stacks made from tissue-mimicking materials, using a wideband laboratory measurement from 2 to 20 GHz. In contrast, \system{} estimates a continuous physiological quantity, site-specific caliper-equivalent skinfold thickness, from in-vivo human measurements using a commodity 460 MHz-bandwidth UWB radar module centered at 7.875 GHz. \system{} therefore differs in target output, sensing substrate, and deployment setting: classification versus continuous regression, phantoms versus human participants, and laboratory broadband characterization versus commodity UWB body-site sensing.

Skinfold measurement estimates subcutaneous fat thickness at standardized anthropometric landmarks using mechanical calipers, and converts multi-site readings to whole-body or regional fat estimates through validated regression equations.
A series of caliper-based algorithms has been developed over five decades for different target populations: Durnin-Womersley \cite{durnin1974body} for clinical research, Jackson-Pollock \cite{jackson1978generalized, jackson1980women} for fitness practice, Slaughter \cite{slaughter1988skinfold} for pediatrics, Yuhasz \cite{yuhasz1974} and Parrillo \cite{parrillo1996} for athletic populations, and ISAK \cite{stewart2011isak} for international kinanthropometric standardization.
Skinfold measurement is widely deployed in field studies because of its low cost, but the procedure is inherently intrusive: the caliper mechanically pinches a fold of skin.
It also requires a trained second operator, introduces significant inter-rater variability, and is limited to subcutaneous fat \cite{machado2025skinfold}.
Consumer-market bioelectrical pincers (``Skinfold Caliper''-style clamp meters) combine these two intrusions, pressing electrodes onto a pinched skinfold and typically requiring a second person for accurate site placement.
We use skinfold calipers as the ground-truth reference in this work.
\system{} replicates their site-specific subcutaneous fat estimates non-intrusively, without pinching the skin and without requiring a trained second operator.

\subsection{Radio Frequency and Radar Sensing for Health} \label{sec:related:radar}

Radio Frequency (RF) sensing has been extensively studied for contactless respiration and heart rate monitoring. Adib et al.'s Vital-Radio \cite{adib2015vitalradio} and subsequent work have demonstrated that FMCW and UWB radar can track thoracic micro-displacements at millimeter resolution. These systems leverage similar range-profile extraction pipelines to \system{}, but target dynamic physiological signals rather than static tissue composition.
Radar-based activity recognition has used Doppler signatures and range-Doppler maps to classify gestures and movements \cite{lien2016soli}. \system{} shares the hardware platform but differs fundamentally in target: we seek to infer structural tissue composition from quasi-static reflections, not dynamic motion.
Medical ground-penetrating radar has been explored for breast tumor detection and bone density estimation \cite{klemm2009radar}. These systems use custom, high-power hardware arrays optimized for clinical settings. \system{} is distinguished by its exclusive reliance on commodity, single-antenna UWB modules and its focus on body-fat estimation as an accessible health metric.
Microwave-based techniques have been proposed for tissue dielectric characterization \cite{gabriel1996dielectric}. Unlike direct-contact impedance methods, UWB radar probes tissue non-invasively at standoff distance. \system{} extends this direction to a consumer-deployable scenario, demonstrating feasibility with hardware that neither pinches the skin nor passes current through the body, and that requires no specialized setup.

\subsection{Health Sensing Using Commodity Wireless and Mobile Devices}
\label{sec:related:ubicomp}

Commodity wireless and mobile devices can be repurposed for unobtrusive physiological measurement:
WiFi channel state information for gait and fall detection \cite{hsu2017wigait, palipana2018falldefi}, smartphone accelerometers for activity and gait analysis \cite{kwapisz2011activity}, and PPG-based wearables for cuffless blood pressure estimation \cite{mukkamala2015cuffless}. These works establish the value of leveraging pervasive sensing infrastructure for health monitoring. \system{} aligns with this trajectory, uniquely positioning UWB, a technology now embedded in consumer smartphones and accessories, as a body-composition radar, extending the frontier of ambient health monitoring to a metric previously confined to clinical settings.

\section{Preliminary}\label{sec:preliminary}

This section provides the technical background for UWB-based body fat measurement.
\Cref{sec:prelim:tissue} describes the layered anatomical structure of human tissue and its corresponding dielectric properties.
\Cref{sec:prelim:cir} derives the relationship between UWB signals and the fat thickness.

\subsection{Layered Human Tissue Model}\label{sec:prelim:tissue}

\begin{figure}[t]
    \centering
    \begin{subfigure}{0.33\linewidth}
        \centering
        \includegraphics[width=\linewidth]{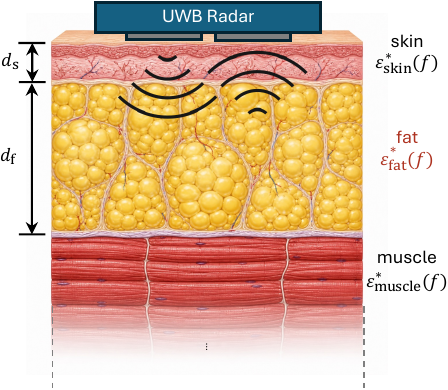}
        \caption{UWB signal and the skin-fat-muscle anatomical structure.}\label{fig:prelim:a}
    \end{subfigure}
    \hfill
    \begin{subfigure}{0.33\linewidth}
        \centering
        \includegraphics[width=\linewidth]{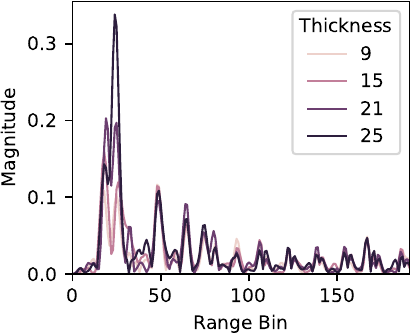}
        \caption{Time-Domain Magnitude.}\label{fig:prelim:b}
    \end{subfigure}
    \hfill
    \begin{subfigure}{0.33\linewidth}
        \centering
        \includegraphics[width=\linewidth]{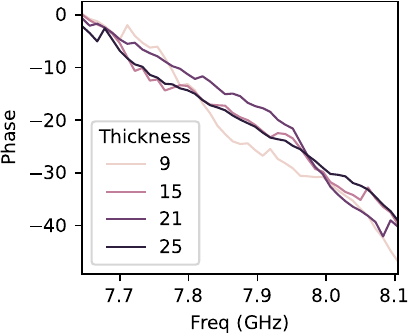}
        \caption{Frequency-Domain Phase.}\label{fig:prelim:c}
    \end{subfigure}
    \caption{Skin-fat-muscle anatomical structure, and UWB signal signatures under different fat thicknesses.}\label{fig:prelim_signal}
\end{figure}

As shown in \Cref{fig:prelim:a}, \system{} measures the fat by placing the UWB radar on the target body site, with the antenna face flush against the skin and the radar boresight oriented normal to the local skin surface.
The UWB radar transmits short pulses into the body. These pulses reflect from tissue interfaces beneath the antenna, and the reflected signals are captured by the receive antenna as complex CIR measurements.
Estimating the fat thickness can be formulated as an inverse problem: given the measured CIR, infer the thicknesses of the underlying tissue layers beneath the antenna. Solving this inverse problem requires a forward model that maps tissue geometry to the received CIR.

We first describe the general skin-fat-muscle anatomical structure used in our model. As shown in \Cref{fig:prelim:a}, the tissue region directly beneath the antenna at each target site can be approximated as a three-layer stack consisting of skin, subcutaneous fat, and muscle~\cite{gasmelseed2025parametric, streeter2024classification}. 
The outermost layer is \emph{skin}, with thickness \(d_s\), comprising the epidermis and dermis with a combined thickness of approximately 1--3~mm and relatively uniform composition.
Below the skin lies \emph{subcutaneous adipose tissue} (SAT), with thickness \(d_f\). SAT is an adipocyte-rich connective tissue layer that stores energy and provides thermal insulation. We define the fat thickness measured in this work as SAT thickness \(d_f\). SAT thickness varies systematically across anatomical sites and depends on factors such as sex and overall adiposity, with reported values of approximately 5--30~mm at clinically relevant skinfold sites, including the abdomen, suprailiac region, triceps, and anterior thigh~\cite{shen2003adipose, jackson1978generalized}.
Beneath the SAT lies skeletal \emph{muscle}, a dense, well-perfused tissue layer that typically extends several centimeters in depth, depending on the anatomical site, before transitioning to deeper structures such as fascia, bone, or visceral compartments.

We model this anatomy as a stratified tissue stack comprising a skin layer of thickness \(d_s\), a fat layer of thickness \(d_f\), and a muscle half-space. Each layer is locally homogeneous and characterized by its frequency-dependent complex permittivity: \(\varepsilon_{\text{skin}}(f)\), \(\varepsilon_{\text{fat}}(f)\), and \(\varepsilon_{\text{muscle}}(f)\).
We assume each layer to be locally homogeneous because the antenna footprint is small, on the order of \(30 \times 5\)~mm. Over this limited area, tissue thickness and composition vary little compared with their variation across standard skinfold sites~\cite{lazebnik2007large}. Thus, within the antenna footprint, the tissue stack can be reasonably approximated as having nearly planar interfaces and locally uniform material properties.
We also note that UWB signals attenuate rapidly after entering muscle tissue, making reflections from deeper structures, such as muscle, bone, and visceral compartments, negligible relative to the system noise floor~\cite{sarestoniemi2020uwb, vculjak2020wireless}. Therefore, the dielectric contrast at the fat-muscle boundary produces a dominant deep interface reflection.

\subsection{CIR Observability of Fat Thickness}\label{sec:prelim:cir}

We next examine how the layered tissue structure appears in the measured CIR. 
A UWB CIR records the aggregate response of propagation paths between the transmitter and receiver, including direct leakage, antenna-body coupling, and reflections from tissue boundaries. 
In our setting, these components are not observed as isolated physical echoes. 
They are mixed by the finite radar bandwidth, the antenna response, and the short propagation distances inside tissue.

A direct time-domain interpretation of the CIR magnitude is insufficient for fat-thickness estimation. 
The fat-muscle reflection is delayed from the skin-fat reflection by the round-trip propagation time through the fat layer,
\begin{equation}
\Delta \tau_f = \frac{2\mathrm{Re}\{n_f\}d_f}{c},
\end{equation}
where \(n_f\) is the complex refractive index of fat and \(c\) is the speed of light in free space. 
Using \(\mathrm{Re}\{n_f\}=3.09\), this delay is \(73\)--\(440\)~ps for \(d_f=5\)--\(30\)~mm. 
The effective receive bandwidth of our radar is \(B=460\)~MHz, which corresponds to a temporal resolution on the order of \(1/B=2.17\)~ns. 
Thus, the skin-fat and fat-muscle responses fall within the same resolution cell and merge in the CIR magnitude. 
As shown in \Cref{fig:prelim:b}, the measured \(|h[n]|\) forms a broad composite response rather than distinct interface peaks. 
This overlap motivates moving from time-domain magnitude to the complex frequency response.

Although the tissue interfaces overlap in the time-domain magnitude, their sub-resolution delays are still encoded in the complex frequency response. 
We therefore transform the measured CIR \(h[n]\) into its in-band frequency response \(H(f)\) and inspect both magnitude and phase over the radar bandwidth. 
This representation is better matched to layered tissue propagation because a small path-length change produces a frequency-dependent phase shift even when the corresponding echoes occupy the same delay bin. 
The relevant material property is the complex permittivity of each tissue. 
Its real component determines the phase velocity through the layer, while its loss component determines attenuation. 
Thus, fat thickness can affect the received signal through two coupled mechanisms: it changes the phase accumulated during round-trip propagation through fat, and it changes the attenuation of the fat-muscle return.

We use tissue dielectric properties to quantify the phase and attenuation effects. 
For each tissue \(n \in \{\mathrm{skin},\mathrm{fat},\mathrm{muscle}\}\), the complex relative permittivity \(\varepsilon_n^*(f)\) and the corresponding complex refractive index \(n_n(f)\) shows in \Cref{eq:epsn}.
\begin{equation}
\label{eq:epsn}
\varepsilon_n^*(f) = \varepsilon_n'(f) - j\frac{\sigma_n(f)}{2\pi f\varepsilon_0}
\qquad
n_n(f)=\sqrt{\varepsilon_n^*(f)}
\end{equation}
where \(\varepsilon_n'(f)\) is the real relative permittivity, \(\sigma_n(f)\) is conductivity, and \(\varepsilon_0\) is the vacuum permittivity. 
The real part of \(n_n(f)\) controls phase accumulation, while the imaginary part controls propagation loss. 
\Cref{tab:prelim-tissue} reports these quantities at the radar center frequency \(f_c=7.875\)~GHz using the IT'IS tissue property database~\cite{hasgall2022itis}. 
The table shows that subcutaneous fat has a refractive index that differs from both skin and muscle, while also having lower attenuation than its neighboring tissues. 
This contrast makes the fat layer observable in principle: it changes the propagation phase of the fat-muscle return, and it does so without fully suppressing the returning component over the target thickness range.

\begin{table}[t]
\centering
\caption{Tissue complex permittivity, refractive index, and one-way attenuation at \(f_c = 7.875\)~GHz from the IT'IS database~\cite{hasgall2022itis}.}\label{tab:prelim-tissue}
\begin{tblr}{colspec = {X[1,l]X[1,c]X[1,c]X[1,c]X[1,c]}, row{1} = {font=\bfseries}}
\toprule
Tissue & \(\varepsilon_r'\) & \( \sigma \)~(S/m) & \(n = \sqrt{\varepsilon_r^*}\) & Atten.~(dB/cm) \\
\midrule
Skin & \(33.3\) & \(5.69\) & \(5.88 - 1.11j\) & \(15.84\) \\
Fat & \(4.77\) & \(0.43\) & \(2.20 -0.23j\) & \(3.23\) \\
Muscle & \(45.7\) & \(7.63\) & \(6.88 - 1.27j\) & \(18.15\) \\
\bottomrule
\end{tblr}
\end{table}

Fat thickness produces a large phase change because the fat-muscle component traverses the fat layer twice. 
Ignoring attenuation for the moment, the approximate phase \(\phi_f(f;d_f)\) accumulated by round-trip propagation through fat and the phase changes \(\Delta \phi_f(f)\) by thickness changes \(\Delta d_f\) are shown in \Cref{eq:phase}.
\begin{equation}
\phi_f(f;d_f) \approx -\frac{4\pi f\,\mathrm{Re}\{n_f\}d_f}{c}
\qquad
\Delta \phi_f(f) \approx -\frac{4\pi f\,\mathrm{Re}\{n_f\}\Delta d_f}{c}
\qquad
\Delta \phi_B(d_f) \approx \frac{4\pi B\,\mathrm{Re}\{n_f\}d_f}{c}
\label{eq:phase}
\end{equation}
Using \(f_c=7.875\)~GHz and \(\mathrm{Re}\{n_f\}=2.20\), a \(1\)~mm change in fat thickness changes the round-trip phase by approximately \(0.73\)~rad, or \(41.6^\circ\). 
This phase change is much larger than the corresponding shift in the time-domain magnitude envelope. 
Thus, two fat thicknesses that cannot be separated as distinct CIR peaks can still produce distinguishable complex frequency responses.

A single-frequency phase measurement is not sufficient because phase wraps with thickness. 
At \(f_c\), the phase period corresponds to \(\lambda_f/2 \approx 8.65\)~mm in fat, where \(\lambda_f=c/(f_c\mathrm{Re}\{n_f\})\). 
The UWB bandwidth reduces this ambiguity because fat thickness also changes the phase slope across frequency. 
Over the effective receive bandwidth \(B=460\)~MHz, the approximate phase variation \(\Delta \phi_B(d_f)\) induced by round-trip propagation through fat is shown in \Cref{eq:phase}.
This gives \(0.42\)~rad at \(d_f=10\)~mm and \(1.27\)~rad at \(d_f=30\)~mm. 
These changes remain below \(\pi\) over the band, so the broadband trend can be inspected without relying on phase unwrapping across the measured support.

\Cref{fig:prelim:c} confirms this frequency-domain intuition in measured data. 
The phase response changes systematically with fat thickness even though the corresponding time-domain magnitude responses overlap. 
This observation provides the empirical counterpart to the propagation analysis: fat information is not exposed as a separate time-domain echo, but it remains present in the complex CIR. 
We therefore retain the in-band complex frequency response for \system{} rather than reducing the measurement to peak location or peak magnitude in \(|h[n]|\).

\section{System Design}\label{sec:system}

\subsection{Overview}\label{sec:system:overview}

\begin{figure}[t]
\centering
\includegraphics[width=\linewidth]{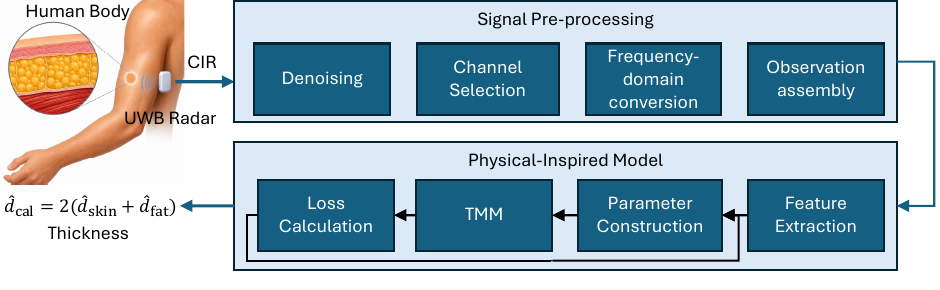}
\caption{System flow of \system{}.} \label{fig:system_overview}
\end{figure}

As shown in Figure~\ref{fig:system_overview}, \system{} consists of three components: 
(1) a commodity UWB radar as the sensing front-end (\Cref{sec:impl:hardware}); 
(2) a deterministic signal processing pipeline that denoises the raw CIR frames and turns them into the in-band channel response (\Cref{sec:system:pipeline});
and (3) a physics-inspired model that infers caliper-equivalent skinfold thickness \(\hat d_{cal} = 2(\hat d_s + \hat d_f)\) from in-band channel response (\Cref{sec:system:model}). 
The system is designed to operate with the radar placed on the target body site, without skin pinching or assistance from a second person.

\subsection{Body-Site-Aware Sensing}\label{sec:system:sites}

The skinfold-caliper algorithms discussed in~\Cref{sec:related:methods} prescribe a small set of anthropometric landmarks designed to capture regional fat depots.
The landmarks are anchored on stable bony or anatomical references, and each algorithm calibrates its regression equation against the specific subset it prescribes.
Different algorithms make different trade-offs: compact protocols sacrifice anatomical coverage for measurement speed, comprehensive protocols sample more landmarks for finer regional discrimination, and pediatric protocols pivot toward the limbs because youth fat distribution differs from adult.
\Cref{tab:skinfold-sites} catalogs the site prescriptions of these algorithms across the standardized anthropometric landmarks they use.
The union of these prescriptions defines the body-composition landscape over which subcutaneous-fat measurement is clinically meaningful, and \system{} must situate its validation within that landscape.

\begin{table}[t]
    \centering
    \caption{Anthropometric sites prescribed by mainstream caliper-based body-fat algorithms, alongside the sites measured by \system{}. A check mark indicates that the algorithm uses the corresponding landmark. The Sup.\ column merges the suprailiac landmark used by Durnin-Womersley, Jackson-Pollock, Yuhasz, and Parrillo with the ISAK iliac-crest and supraspinale landmarks, which all lie within a few centimeters of the anterior superior iliac spine.}\label{tab:skinfold-sites}
    \begin{tblr}{
      colspec = {X[5.3,l,m] *{9}{X[1,c,m]}},
      row{1} = {font=\bfseries},
      row{10} = {font=\bfseries, bg=gray!15}
    }
    \toprule
    Algorithm & Chest & Tri. & Biceps & Sbs. & MidAx. & Abd. & Sup. & Thigh & Calf \\
    \midrule
    Durnin-Womersley 4-site \cite{durnin1974body} & & \checkmark & \checkmark & \checkmark & & & \checkmark & & \\
    Jackson-Pollock 3-site (M) \cite{jackson1978generalized} & \checkmark & & & & & \checkmark & & \checkmark & \\
    Jackson-Pollock 3-site (F) \cite{jackson1980women} & & \checkmark & & & & & \checkmark & \checkmark & \\
    Jackson-Pollock 7-site \cite{jackson1978generalized} & \checkmark & \checkmark & & \checkmark & \checkmark & \checkmark & \checkmark & \checkmark & \\
    Slaughter (youth) \cite{slaughter1988skinfold} & & \checkmark & & \checkmark & & & & & \checkmark \\
    Yuhasz 6-site \cite{yuhasz1974} & & \checkmark & & \checkmark & & \checkmark & \checkmark & \checkmark & \checkmark \\
    Parrillo 9-site \cite{parrillo1996} & \checkmark & \checkmark & \checkmark & \checkmark & & \checkmark & \checkmark & \checkmark & \checkmark \\
    ISAK 8-site \cite{stewart2011isak} & & \checkmark & \checkmark & \checkmark & & \checkmark & \checkmark & \checkmark & \checkmark \\
    \midrule
    \system{} (this work) & \checkmark & \checkmark & & & & \checkmark & \checkmark & \checkmark & \\
    \bottomrule
    \SetCell[c=10]{l} {\footnotesize Note. Tri.: Triceps; Sbs.: Subscapular; MidAx.: Midaxillary; Abd.: Abdomen; Sup.: Suprailiac}
    \end{tblr}
\end{table}

\system{} measures fat thickness at five sites: the chest, the triceps, the abdomen, the suprailiac, and the thigh.
These five sites are exactly the union of the Jackson-Pollock 3-site (M) and 3-site (F) protocols (Table~\ref{tab:skinfold-sites}), which are the most widely used skinfold-caliper algorithms in commercial body-fat estimation and consumer fitness practice~\cite{jackson1978generalized, jackson1980women}.
Covering this union lets \system{} be validated end-to-end against either Jackson-Pollock equation regardless of subject sex, and concentrates the validation evidence on the family of skinfold equations consumers and practitioners encounter most often.
Each landmark is also reachable by a single seated or standing user without a second operator, so the validation protocol matches the deployment protocol.

Each site contributes a distinct anatomical context to the validation panel.
The chest site sits as a diagonal fold midway between the anterior axillary line and the nipple, representing the truncal-upper context with thin subcutaneous fat over the pectoralis major.
The triceps site sits on the posterior aspect of the upper arm at the midpoint between the acromion and the olecranon, representing the upper-limb context with thin-to-moderate fat over a curved muscular substrate.
The abdomen site sits 2--3~cm lateral to the umbilicus, representing the truncal-anterior context with the largest fat-thickness range in the body over a soft underlying muscle wall.
Abdominal subcutaneous thickness correlates with visceral adiposity in adults~\cite{shen2003adipose}.
The suprailiac site sits at the anterior superior iliac spine, representing the truncal-lateral context with a bony substrate beneath thin-to-moderate fat.
The thigh site sits midway between the inguinal crease and the patella on the anterior aspect, representing the lower-limb context with moderate-to-thick fat over a thick muscle base.
Gluteofemoral subcutaneous fat carries a comparatively lower metabolic risk profile than truncal fat~\cite{manolopoulos2010gluteofemoral}.
Validating across this panel produces per-site error breakdowns that reveal which anatomical contexts dominate the residuals rather than absorbing them into a single global metric.

\subsection{Signal Processing Pipeline}\label{sec:system:pipeline}
The six-stage signal processing pipeline turns raw CIR frames into the in-band channel response \(H(f)\) that feeds the physics-inspired model in \Cref{sec:system:model}.

\paragraph{Step 1: CIR acquisition.}
We acquire one recording per measurement using the parameters described in \Cref{sec:impl:hardware}.
Each recording contains \(M\) frames over the four \(2\,\mathrm{TX}\times 2\,\mathrm{RX}\) channels, with \(192\) complex range bins per frame.

\paragraph{Step 2: Denoising.}
Three operations suppress noise and motion artifacts inside each channel before downstream processing.
We apply a sliding-window average across consecutive frames with window length \(W=100\). This operation reduces frame-level noise and produces one denoised CIR per window. Because standing posture, respiration, and small hand-device motion can rotate the complex phase across frames, we do not interpret this step as ideal coherent integration. The ideal \(10\log_{10}W\) gain is only an upper bound for perfectly phase-aligned samples.

A recording therefore yields one denoised CIR per window, and a single recording produces multiple denoised CIRs across the available windows.
The stage output is one denoised CIR \(h^{(c)}[n]\) per channel \(c\) per window.

\begin{figure}[t]
    \centering
    \begin{subfigure}{0.245\linewidth}
        \centering
        \includegraphics[width=\linewidth]{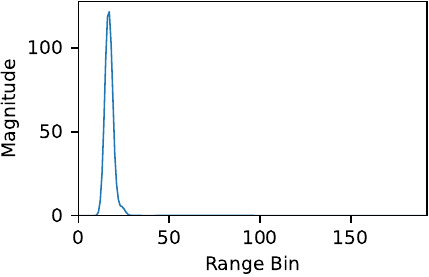}
        \caption{TX0, RX0.}\label{fig:direct_leak:00}
    \end{subfigure}
    \begin{subfigure}{0.245\linewidth}
        \centering
        \includegraphics[width=\linewidth]{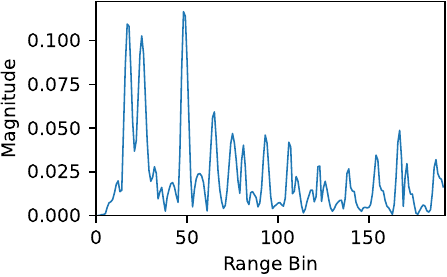}
        \caption{TX0, RX1.}\label{fig:direct_leak:01}
    \end{subfigure}
    \begin{subfigure}{0.245\linewidth}
        \centering
        \includegraphics[width=\linewidth]{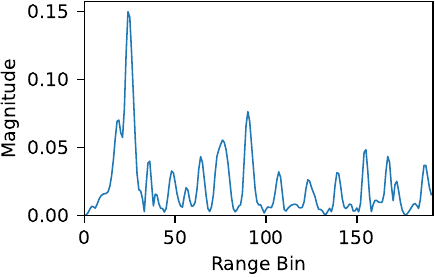}
        \caption{TX1, RX0.}\label{fig:direct_leak:10}
    \end{subfigure}
    \begin{subfigure}{0.245\linewidth}
        \centering
        \includegraphics[width=\linewidth]{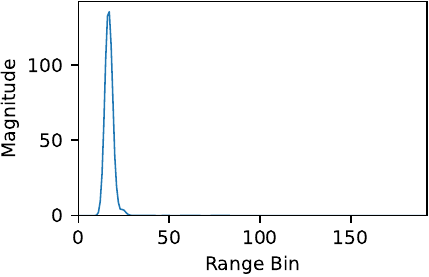}
        \caption{TX1, RX1.}\label{fig:direct_leak:11}
    \end{subfigure}
    \caption{Per-channel CIR magnitude on the X7F202.
    The self-channels \((\mathrm{TX0},\mathrm{RX0})\) and \((\mathrm{TX1},\mathrm{RX1})\) exhibit a saturating direct-leak peak at the body bin, while the cross-channels \((\mathrm{TX0},\mathrm{RX1})\) and \((\mathrm{TX1},\mathrm{RX0})\) preserve the body return.}\label{fig:direct_leak}
\end{figure}

\paragraph{Step 3: Channel selection.}
The X7F202 exposes four \(\mathrm{TX}\times\mathrm{RX}\) channels, but they do not carry equivalent information.
The two self-channels \((\mathrm{TX0},\mathrm{RX0})\) and \((\mathrm{TX1},\mathrm{RX1})\) suffer a critical direct-leak problem.
TX-to-RX coupling on the same antenna pair sits tens of dB above the body return and saturates the front-end at the body bin (\Cref{fig:direct_leak}).
The two cross-channels \((\mathrm{TX0},\mathrm{RX1})\) and \((\mathrm{TX1},\mathrm{RX0})\) avoid this leakage path because the transmit and receive antennas are physically separated.
We therefore retain only the two cross-channels and discard the self-channels for the remainder of the pipeline.

\paragraph{Step 4: Frequency-domain conversion.}
We zero-pad each retained CIR and apply the FFT to obtain the channel response \(H^{(c)}(f)\) on the receiver grid.
The native frequency resolution is \(F_s/N \approx 10.94\)~MHz.

\paragraph{Step 5: Observation assembly.}
We restrict \(H^{(c)}(f)\) to the receiver passband, keeping 43 in-band complex bins per channel.
We stack the two cross-channel responses into one \emph{observation} \(H(f) = [\,H^{(0,1)}(f),\;H^{(1,0)}(f)\,]\), the unit consumed by the physics-inspired model.
Because each observation is built from one sliding-window average over \(W\) consecutive frames, a single recording yields multiple observations.

\paragraph{Step 6: Physics-inspired model inference.}
Each observation \(H(f)\) is fed to the physics-inspired model in \Cref{sec:system:model}.
The model internally parameterizes skin thickness \(\hat d_s\) and SAT thickness \(\hat d_f\), output \(\hat d_{cal} = 2(\hat d_s+\hat d_f)\).
It also returns a per-observation residual \(\sum_{c,f}\|\hat H^{(c)}(f) - H^{(c)}(f)\|^2\) that serves as a confidence signal.

\subsection{Physics-Inspired Model}\label{sec:system:model}

\begin{figure}[t]
\centering
\includegraphics[width=\linewidth]{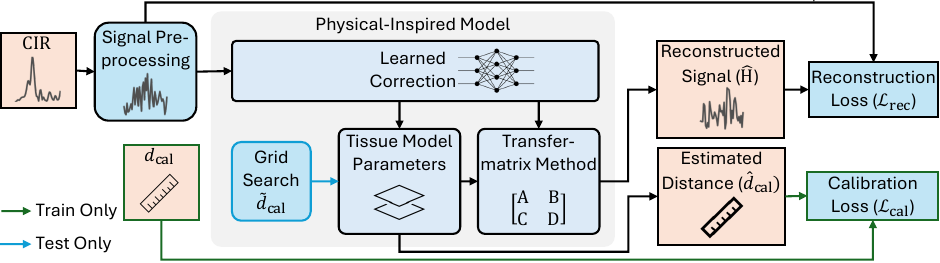}
\caption{Physics-inspired forward model of \system{}.}\label{fig:system_model}
\end{figure}

The physics-inspired forward map provides a structured feature space; learned corrections compensate for unmodeled device and tissue residuals; the headline output is the skinfold thickness \(2(\hat d_s+\hat d_f)\) jointly fit to caliper labels.
The model uses the four-medium transfer-matrix forward map from \Cref{sec:prelim:cir} as its analytic trunk. 
A signal encoder and bounded MLP corrections account for device response, coupling residuals, and tissue-property deviations that are not captured by the nominal layered model.
\Cref{fig:system_model} illustrates the structure.

During training, the model consumes one observation, defined in \Cref{sec:system:pipeline} as the two in-band cross-channel responses, \(H^{(c)}(f) = \operatorname{FFT}(h^{(c)}[n])\), 
where \(c \in \{(0,1),(1,0)\}\) represent the channel of CIR, \(B\) batches observations.
The caliper ground truth is the skinfold reading at the recording's site, \(d_{\mathrm{cal}} = 2\,(d_s + d_f) \), which is shared across all observations within one recording and couples the predicted skin and fat thicknesses to the operator's measurement.

\paragraph{Tissue Model Parameters.}
Each tissue \(n \in \{\text{air},\text{skin},\text{fat},\text{muscle}\}\) carries a frequency-dependent complex permittivity \(\varepsilon^*_n(f)\) drawn from the IT'IS database \cite{hasgall2022itis}.
Then we calculate wavenumber \( k_n(f) = \frac{2\pi f}{c}\sqrt{\varepsilon^*_n(f)} \) and wave impedance \( \eta_n(f) = \frac{\eta_0}{\sqrt{\varepsilon^*_n(f)}} \) per layer. 
With \( k_n(f) \) and \( \eta_n(f) \), Fresnel reflection coefficient \(\Gamma_n(f)\) at the interface between layers \(n\) and \(n{+}1\), companion transmission \( T_n(f) \), and 
the single-pass propagation \( P_n(f) \) through layer \(n\) of thickness \(d_n\) shows in \Cref{eq:GTP}. 
\begin{equation}
\label{eq:GTP}
\Gamma_n(f) = \frac{\eta_{n+1}(f) - \eta_n(f)}{\eta_{n+1}(f) + \eta_n(f)}, 
\quad 
T_n(f) = 1 + \Gamma_n(f),
\quad
P_n(f) = e^{-jk_n(f)d_n}.
\end{equation}

\paragraph{Transfer-Matrix Method} We build the Fresnel matrix \( M_I^{n} \) for each interface, propagation matrix \( M_P^{n} \) for each layer, and the ordered product over the air-skin-fat-muscle stack and the multilayer reflection at the antenna face in \Cref{eq:FPM}.
\begin{equation}
\label{eq:FPM}
M_I^{n\{n+1\}} = \frac{1}{T_n}\begin{bmatrix} 1 & \Gamma_n \\ \Gamma_n & 1 \end{bmatrix},
\quad
M_P^{n} = \begin{bmatrix} P_n & 0 \\ 0 & P_n^{-1} 
\end{bmatrix},
\quad
M = M_P^{0}\ M_I^{01}\ M_P^{1}\ M_I^{12}\ M_P^{2}\ M_I^{23} = \begin{bmatrix} M_{11} & M_{12} \\ M_{21} & M_{22} \end{bmatrix}.
\end{equation}

\paragraph{Reconstructed Signal} With transfer matrix, we have RF reflection function \( \Gamma^{(0)}(f) \) for multi-layered tissue structures. We model the body-free system response \( P_{\mathrm{sys}}(f) \) as a smooth complex frequency response over the effective passband. Its magnitude is approximated by a Gaussian envelope, and its phase is represented by a second-order polynomial around the center frequency, shows in \Cref{eq:psys}
\begin{equation}
\label{eq:psys}
\Gamma^{(0)}(f) = \frac{M_{21}(f)}{M_{11}(f)},
\qquad
P_{\mathrm{sys}}(f) = A_p e^{-\!\left(\frac{f-f_c}{B_p}\right)^{\!2}}
e^{j\!\left(\phi_0 + 2\pi\tau_p(f-f_c) + \pi a_p(f-f_c)^2\right)}.
\end{equation}
The predicted in-band channel response is the product of the corrected system pulse and the corrected multilayer reflection:
\begin{equation}
\hat H^{(c)}(f) = P_{\mathrm{sys}}^{(c)}(f) \cdot \Gamma^{(0)}(f).
\end{equation}

\paragraph{Learned Correction}
We construct a Signal Encoder \( F_{\mathrm{CIR}} = \mathrm{Encoder}\big(H(f)\big) \in \mathbb{R}^{[B,\,64]}\) using Conv1D stack along the frequency axis projects each observation into a shared latent that conditions the per-observation parameters \(\{g,\,d_s,\,d_f\}\).
The encoder acts on the real-imaginary concatenation \([\Re H,\,\Im H]\) along the frequency axis, projecting into a 64-dimensional latent.
The correction networks predict bounded per-observation residuals conditioned on the encoded signal on \(\varepsilon^*, \Gamma, P_\text{sys}, \Gamma^{(0)}, \text{and} \mathrm{sys}\), clipped to physically plausible bounds \([a,b]\)
The configuration of each MLP matches the dimensionality of its input.
The total trainable count is on the order of \(10^4\), small enough to fit per subject in seconds and lightweight enough to leave the analytic transfer-matrix map dominant in the forward graph.

\paragraph{Loss Functions.}
The training objective combines frequency-domain reconstruction with caliper consistency, as shown in \Cref{eq:loss}.
\begin{equation}
\label{eq:loss}
\mathcal{L}_{\mathrm{total}} = \sum_{b,c,f}|\hat H_b^{(c)}(f)-H_b^{(c)}(f)|_2^2
+
\sum_b (d_{\mathrm{cal},b} - 2(\hat d_{s,b} + \hat d_{f,b}))^2.
\end{equation}
The reconstruction term forces the predicted spectrum to match the measured complex response.
The caliper term anchors the recovered skin and fat thicknesses to the operator's reading.

\paragraph{Testing Procedure}
At test time, \system{} processes each observation independently.
For an observed complex frequency response \(\mathbf{y}\), we enumerate candidate fat and skin thicknesses, \(d_{\mathrm{fat}}\) and \(d_{\mathrm{skin}}\), over a grid constrained by the valid physical range of the tissue model.
For each candidate pair, the learned correction network predicts the correction parameters used by the forward model.
The corrected forward model then reconstructs the expected radar response \(\hat{\mathbf{y}}(d_{\mathrm{fat}}, d_{\mathrm{skin}})\).
We compute the reconstruction loss between this response and the observation.
The predicted thicknesses are the candidate values that produce the smallest loss:
\begin{equation}
(\hat{d}_{\mathrm{fat}}, \hat{d}_{\mathrm{skin}}) = \arg\min_{d_{\mathrm{fat}}, d_{\mathrm{skin}}}\left\| \hat{\mathbf{y}}(d_{\mathrm{fat}}, d_{\mathrm{skin}}) - \mathbf{y} \right\|_2^2 .
\end{equation}
This inference procedure restricts the search to physically plausible tissue configurations and selects the configuration whose reconstructed signal best matches the measured response.

\section{Implementation}
\label{sec:implementation}

\subsection{Hardware Platform}
\label{sec:impl:hardware}

We build the system around the Novelda X7F202 development kit \cite{novelda_x7f202_datasheet}, a commodity UWB impulse-radar transceiver shipped in a small module with the X7F202 SoC and an integrated 2~TX, 2~RX antenna array. 
The X7F202 operates in the same FCC Part 15.503 UWB band as the UWB radios integrated into current smartphones, smartwatches, and item trackers, with comparable RF bandwidth and pulse characteristics. 
The key difference is the firmware-level output. 
Consumer UWB chips usually expose IEEE 802.15.4z ranging outputs or restrict raw signal access through proprietary APIs, whereas the X7F202 exposes the raw complex baseband CIR that \system{} requires.

The device transmits sub-nanosecond pulses centered at \(f_c = 7.875\)~GHz with a TX \(-10\)~dB bandwidth of 750~MHz. 
It digitizes the receive chain at 2100~MS/s through a \(-3\)~dB band of 460~MHz, so the effective processing bandwidth is receiver-limited at \(B \approx 460\)~MHz. 
Each radar frame contains 192 range bins of width \(c\,\Delta t / 2 \approx 7.14\)~cm in air, scaling to about 3.20~cm inside fat. 
The two transmit and two receive antennas yield four channel combinations. 

The module connects to a Raspberry Pi~5, which serves as the embedded host. The Pi runs the Novelda BA22 firmware loader and the Python acquisition flow. The Pi is in turn connected to a workstation laptop over Ethernet for storage and offline analysis. The workstation does not need to be present during a measurement: the Pi stores recordings locally and syncs to the laptop on demand. The module is hand-held by the operator and placed on the target body site. 
There is no fixture or stand.

\subsection{Software Stack}
\label{sec:impl:software}

The acquisition layer runs in Python on the Raspberry Pi using the Novelda RadarDirect callback flow, which delivers complex baseband I/Q frames per channel combination directly to user code. Each recording is written as a single HDF5 file containing one dataset per (chip, TX, RX) tuple, plus the radar configuration and per-frame metadata. 
We collect 1000 frames per recording at 100 fps, giving a 10 s acquisition window. The frames are later divided into short temporal windows for denoising and observation assembly.
We use the \texttt{pyx7configuration} library to generate a chip configuration that satisfies the X7's PRF and timing constraints, with the user-level parameters listed in Table~\ref{tab:impl-config}.

The pre-processing layer runs on the workstation in Python with NumPy and SciPy. It loads the HDF5 recording, applies the steps of Section~\ref{sec:system:pipeline} in order, and produces a sequence of observations $H_{\mathrm{obs}}(f)$ on the 460~MHz RX grid. The physics-inspired model of Section~\ref{sec:system:model} is implemented in PyTorch. The TMM stack matrix, the Gaussian-times-quadratic-phase $P_{\mathrm{sys}}(f)$, and the coupling residual are all written as differentiable forward functions. TMM parameters are PyTorch \texttt{nn.Parameter} objects with explicit physical bounds enforced by softplus or sigmoid reparameterizations. Training uses Adam as optimizer. Inference at deployment is offline on the recorded HDF5 files and converges in under one second per observation on a CPU.

\begin{table}[t]
\centering
\caption{X7F202 chip-level operating configuration used in all recordings, generated by \texttt{pyx7configuration} from the user-level parameters and serialized in \texttt{config.json}.}
\label{tab:impl-config}
\small
\begin{tabular}{ll}
\hline
Parameter & Value \\
\hline
FPS & 100 \\
Frames per recording & 1000 \\
PulsePeriod & 12 ($\rightarrow$ PRF $\approx 10.94$~MHz) \\
MframesPerPulse & 12 ($\rightarrow$ 192 range bins, max range $\approx 13.7$~m in air) \\
PulsesPerIteration & 35 \\
IterationsPerFrame & 1024 \\
TxPower & 4 (maximum) \\
InterleavedFrames & 5 (dual-TX alternating) \\
TxChannelSequence & $\{0, 1\}$ (TX0 and TX1 alternating) \\
RxMaskSequence & $\{3, 3\}$ (both RX active in each subframe) \\
DCRemoval & off (raw I/Q retained) \\
\hline
\end{tabular}
\end{table}

\subsection{Ground Truth Collection}
\label{sec:impl:groundtruth}

We use skinfold caliper measurements as the ground-truth reference for subcutaneous tissue thickness at every body site. A trained operator measures each site three times at the same anatomical landmarks used by the Jackson-Pollock protocol \cite{jackson1978generalized, jackson1980women}, with the mean of the three readings used as the per-site ground truth. Caliper measurements are conducted immediately before each \system{} recording at the same landmark, to minimize biological variation between the two readings. Each \system{} recording is captured with the radar in light skin contact at the same landmark and with no intentional standoff, so that radar and caliper readings sample the same anatomical column. The site list matches the body sites used by the radar protocol in Section~\ref{sec:system:sites}: chest, triceps, abdomen, suprailiac, and thigh.

\begin{figure}[t]
    \centering
    \begin{subfigure}{0.195\textwidth}
        \includegraphics[width=\linewidth]{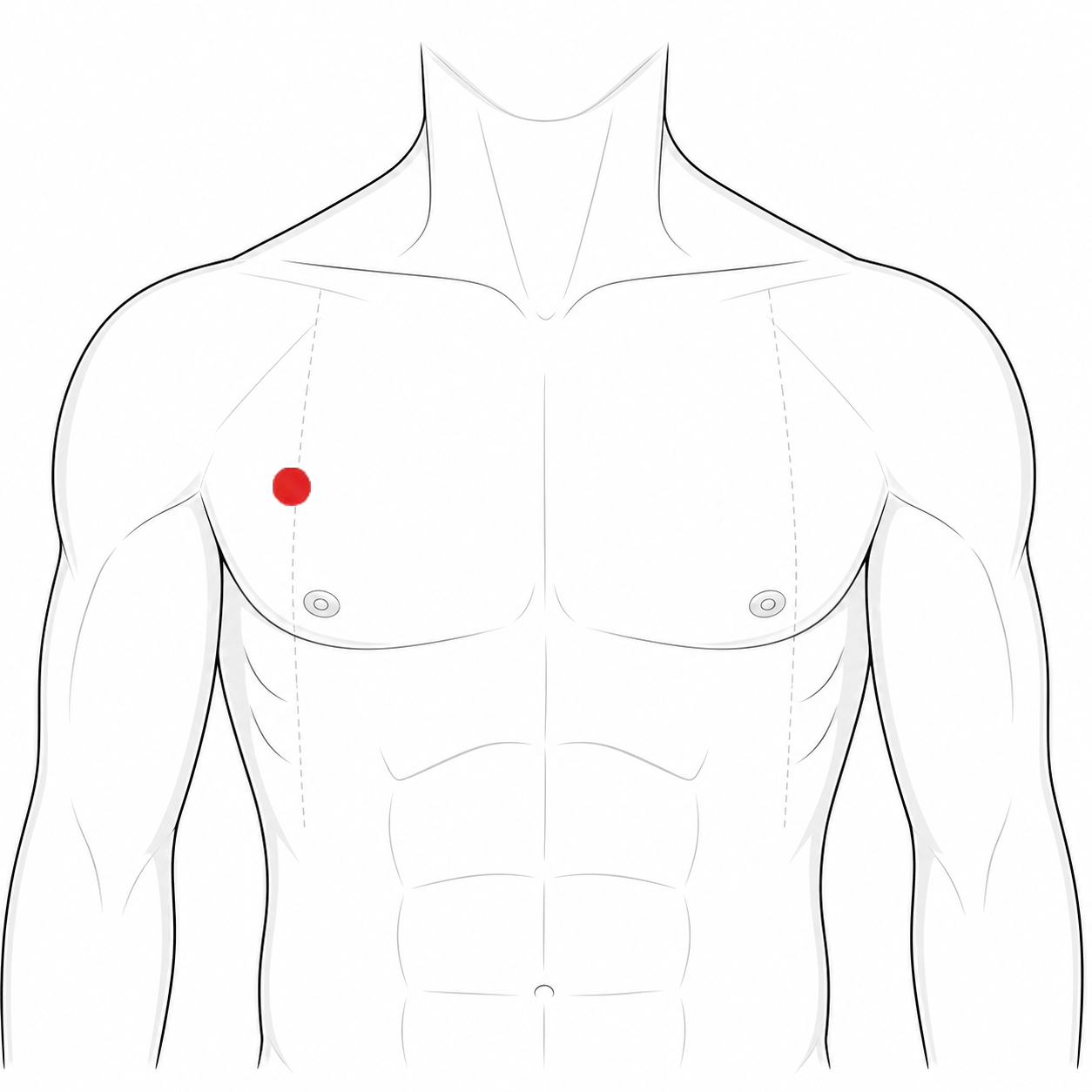}
        \caption{Chest}
    \end{subfigure}
    \hfill
    \begin{subfigure}{0.195\textwidth}
        \includegraphics[width=\linewidth]{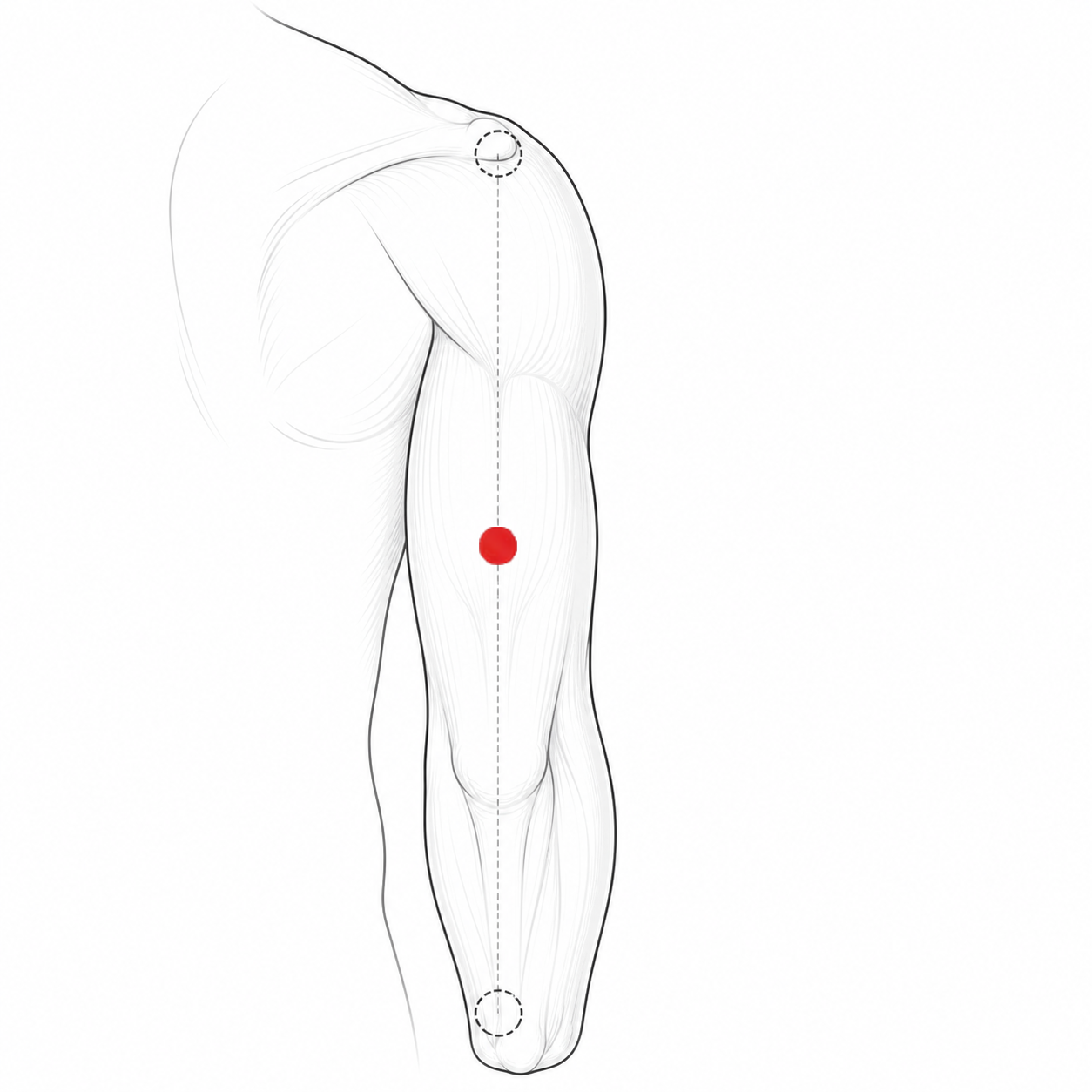}
        \caption{Triceps}
    \end{subfigure}
    \hfill
    \begin{subfigure}{0.195\textwidth}
        \includegraphics[width=\linewidth]{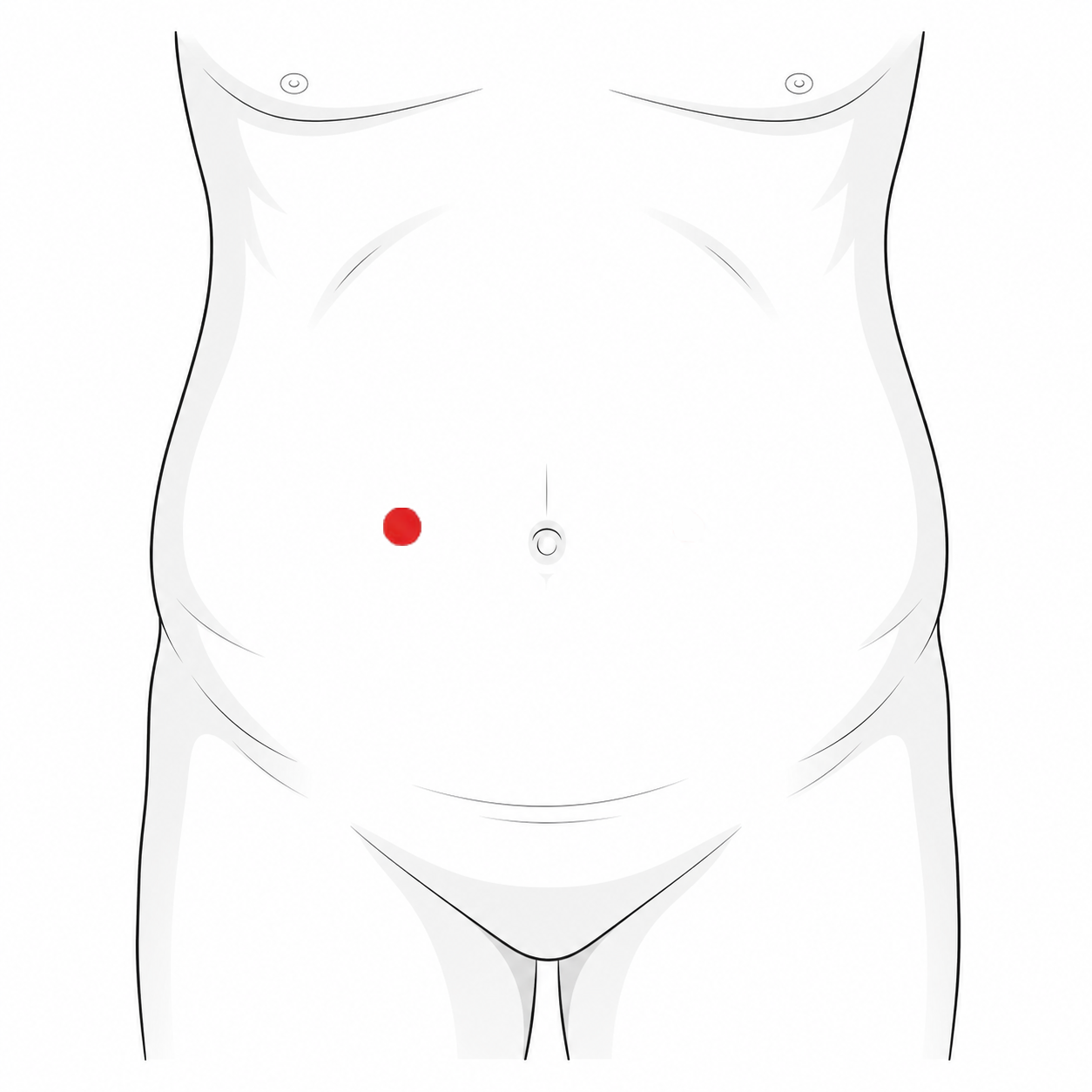}
        \caption{Abdomen}
    \end{subfigure}
    \hfill
    \begin{subfigure}{0.195\textwidth}
        \includegraphics[width=\linewidth]{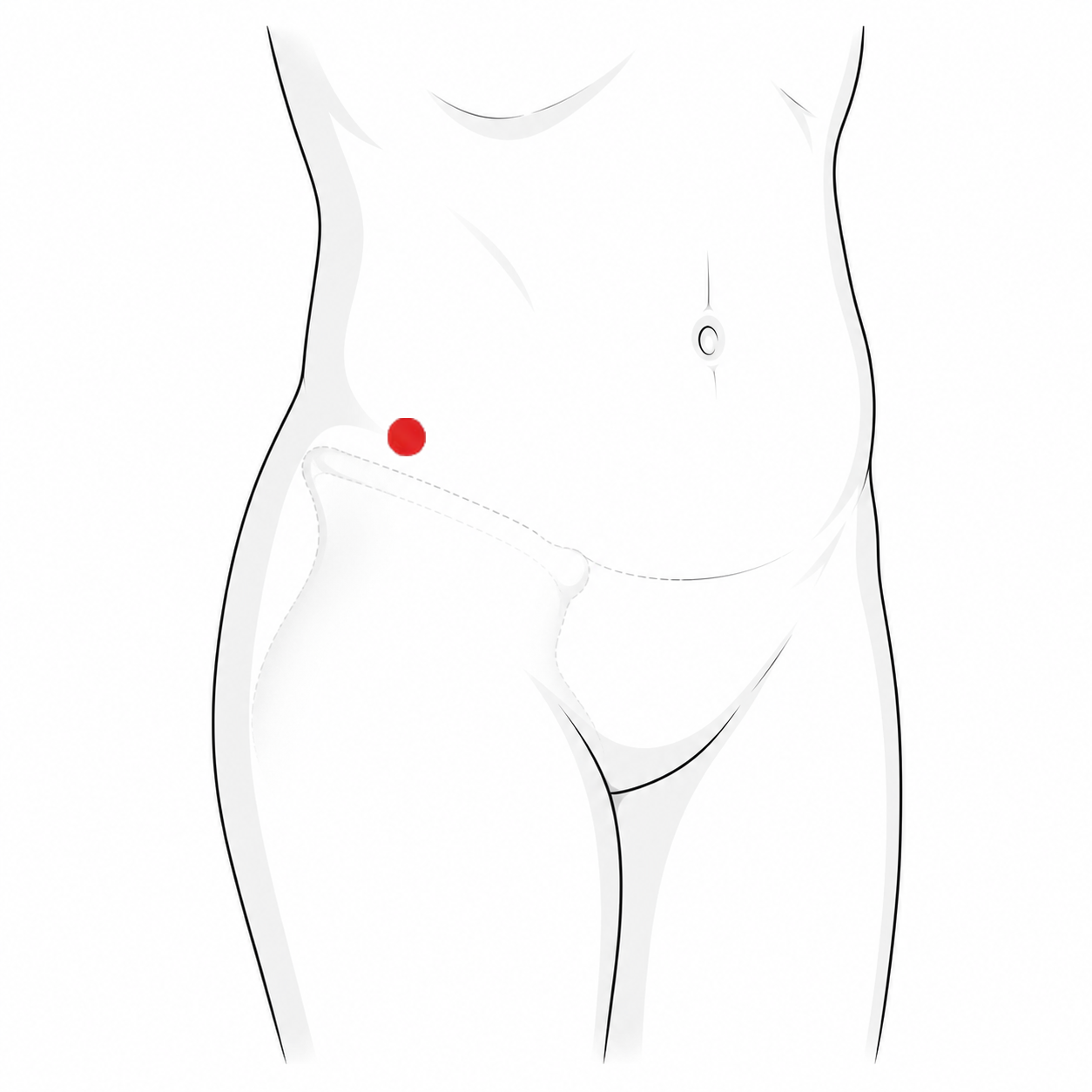}
        \caption{Suprailiac}
    \end{subfigure}
    \hfill
    \begin{subfigure}{0.195\textwidth}
        \includegraphics[width=\linewidth]{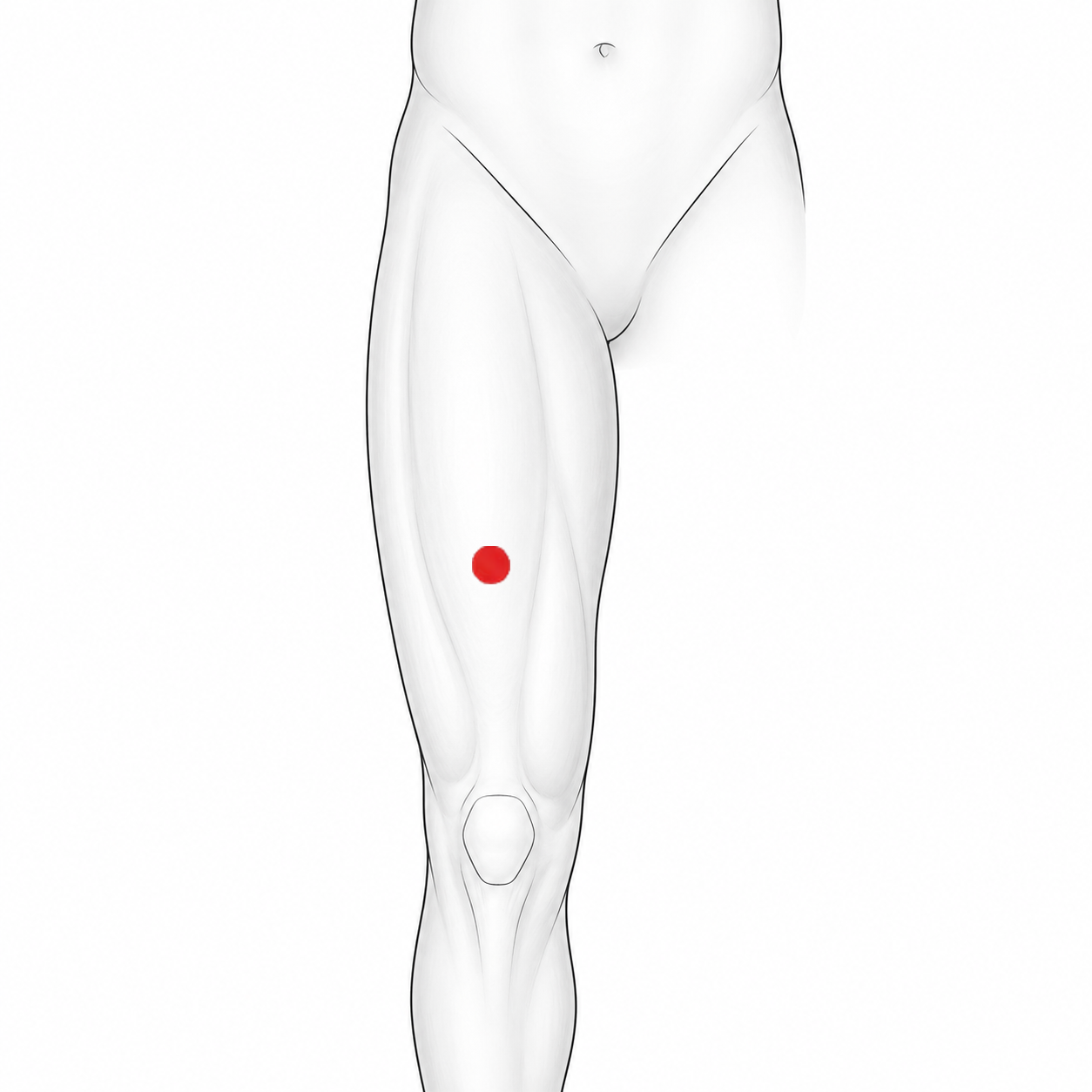}
        \caption{Thigh}
    \end{subfigure}
    \caption{Body-site coverage: the protocol cycles through the five anatomical sites of Section~\ref{sec:system:sites}.}\label{fig:body_sites} \end{figure}

The caliper jaw pinches a skinfold containing two layers of skin and subcutaneous fat. The reading therefore corresponds directly to $y_{\mathrm{caliper}} = 2(d_s + d_f)$, the doubled per-side thickness in millimeters. We do not halve, average, or convert this value; the physics-inspired model in Section~\ref{sec:system:model} is fit to match $y_{\mathrm{caliper}}$ directly through the label-consistency term $\big(y_{\mathrm{caliper}} - 2(d_s + d_f)\big)^2$. The caliper reading is the only ground truth we use, and we do not derive whole-body fat percentage from it. Our system replaces the caliper as a site-specific subcutaneous tissue thickness probe; it does not replace whole-body composition tools such as BIA scales or DEXA.

\subsection{Study Protocol}\label{sec:impl:protocol}

\begin{figure}[t]
\centering
    \includegraphics[width=0.5\linewidth]{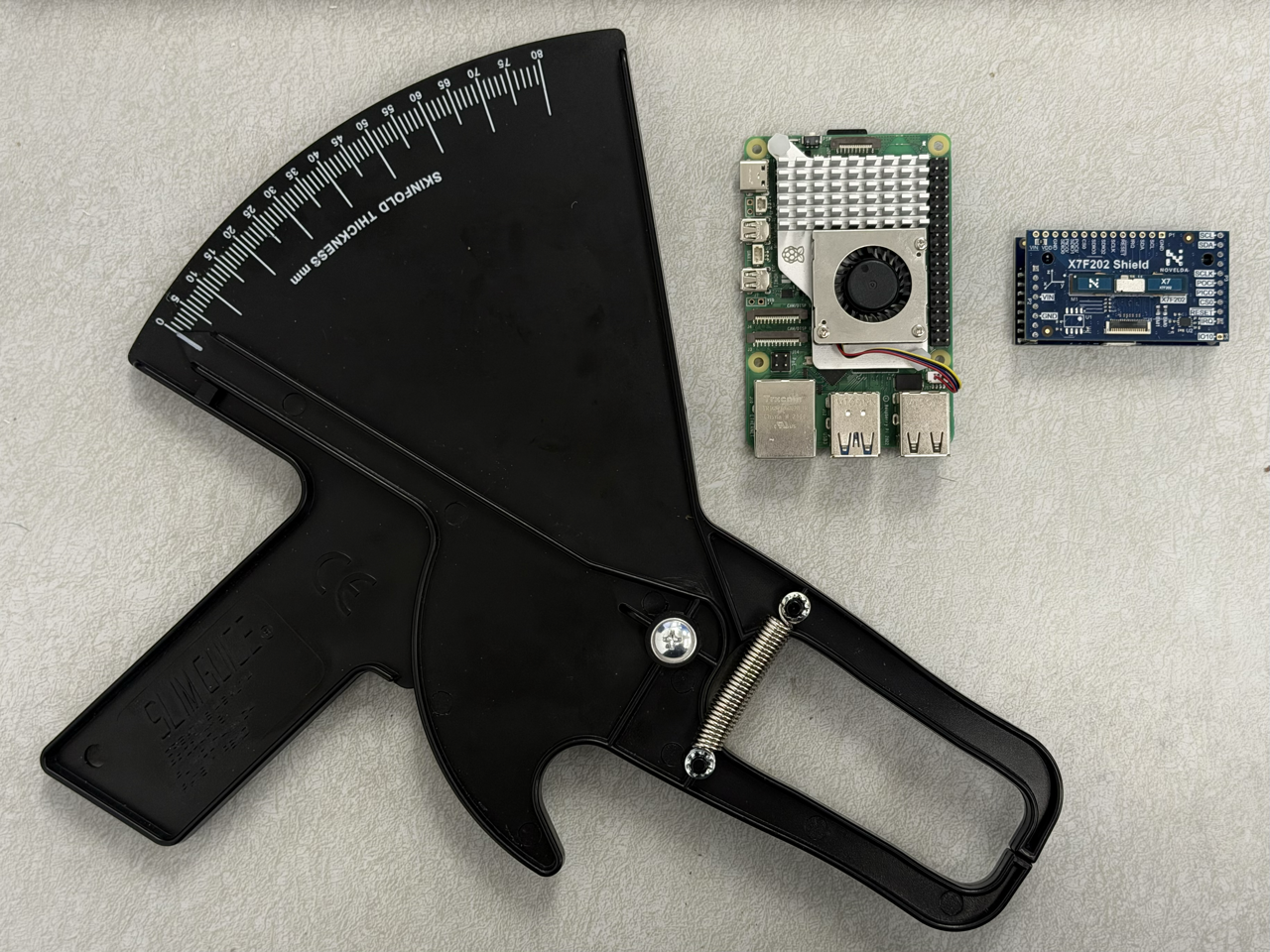}
    \caption{Hardware: the Novelda X7F202 development kit module tethered to a Raspberry Pi 5 host. Groundtruth collected by skinfold caliper.}
\end{figure}

The participant is standing at the appropriate posture for the target site (Section~\ref{sec:system:sites}). 
The target site is exposed and wiped dry. 
The caliper measurement is taken three times. 
The radar recording is then captured by placing the radar on the same landmark for the 10~s acquisition window, following the protocol of Section~\ref{sec:impl:groundtruth}. 
The procedure is repeated 6 times for each site in a randomized order. 
Total session time per participant is approximately 20–25 minutes.

\section{Evaluation}\label{sec:evaluation}

\subsection{Dataset}\label{sec:eval:dataset}

We recruited 15 adult volunteers following the protocol described in \Cref{sec:impl:protocol}. \Cref{tab:eval-demographics} summarizes the cohort. Each participant completed one main accuracy session on bare skin at the five body sites listed in \Cref{sec:system:sites}, and a subset additionally completed the seven sensing-factor ablations of \Cref{sec:eval:ablation}. The total dataset contains 450 paired recordings of radar data and skinfold caliper measurements, which contains 450,000 CIR frames in total, preprocessed in 4,500 observations for model input.

\begin{table}[t]
\centering
\caption{Participant demographics.}\label{tab:eval-demographics}
\begin{tblr}{colspec = {X[0.1,c,m] X[1.2,c,m] X[3,c,m]*{2}{X[1,c,m]} X[3.3,c,m]*{2}{X[1,c,m]} X[3.3,c,m]*{2}{X[1,c,m]} X[3,c,m]*{2}{X[1,c,m]}}}
\toprule
\SetCell[r=2]{}\( N \) & Gender & \SetCell[c=3]{}Age &  &  & \SetCell[c=3]{}Height &  &  & \SetCell[c=3]{}Weight &  &  & \SetCell[c=3]{}BMI & & \\
\cmidrule[lr]{3-5}\cmidrule[lr]{6-8}\cmidrule[lr]{9-11}\cmidrule[lr]{12-14}
 & M/F & \(\mu \pm \sigma \) & \( \min \) & \( \max \) & \(\mu \pm \sigma \) & \( \min \) & \( \max \) & \(\mu \pm \sigma \) & \( \min \) & \( \max \) & \(\mu \pm \sigma \) & \( \min \) & \( \max \) \\
\midrule
15 & 8/7 & 26.1 \(\pm \) 5.3 & 21 & 44 & 171.8 \(\pm \) 6.9 & 160 & 182 & 67.9 \(\pm \) 11.2 & 51 & 92 & 22.9 \(\pm \) 2.8 & 18.8 & 29.1 \\
\bottomrule
\end{tblr}
\end{table}

\subsection{Train/Test Split}\label{sec:eval:split}
We use two evaluation protocols. The primary accuracy evaluation uses leave-one-subject-out (LOSO) cross-validation over the 15 participants. In each fold, the model is trained only on the default-protocol recordings from 14 participants. The held-out participant contributes no data during training. 
All predictions for that participant are produced at test time and are then pooled across the 15 folds to compute the metrics in \Cref{sec:eval:quantitative}.
During training and testing, the model is not given the anatomical site label of an input observation.
Thus, each prediction is inferred from the radar response alone, rather than from site-specific model selection or site-conditioned parameters.

The sensing-factor ablation uses a separate condition-holdout protocol. The model is trained only on default-protocol recordings. Recordings collected under the seven perturbed conditions (A1--A7) are excluded from training and used only for robustness evaluation in \Cref{sec:eval:ablation}. Thus, the training objective never uses labeled examples collected with standoff, clothing, lotion, sweat, off-normal incidence, alternative environments, or pedestrian interference.

\subsection{Metrics}\label{sec:eval:metrics}

The model in \Cref{sec:system:model} predicts skin thickness $\hat d_s$ and fat thickness $\hat d_f$ at each site. The caliper jaw measures a fold of two skin-fat layers, so the matching ground truth is $y_{\mathrm{caliper}} = 2(d_s + d_f)$ in mm. We therefore evaluate accuracy by comparing $2(\hat d_s + \hat d_f)$ to the caliper reading at each site, in millimeters of skinfold thickness. We report mean absolute error (MAE), root mean square error (RMSE), and \(R^2\). We additionally report Bland-Altman analysis \cite{bland1986statistical} to characterize agreement and systematic bias. 
\subsection{Quantitative Analysis}\label{sec:eval:quantitative}

We evaluate \system{} along three axes: pooled accuracy across the cohort, ablation studies, and body-fat calculation.

\begin{figure}[t]
\centering
\begin{subfigure}{0.41\textwidth}
    \includegraphics[width=\linewidth]{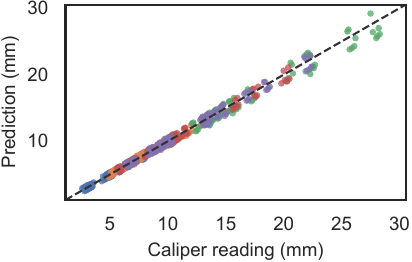}
    \caption{Scatter of \system{} predictions $2(\hat d_s + \hat d_f)$ against caliper reading $y_{\mathrm{caliper}}$. The dashed line is the $y = x$ reference.}
\end{subfigure}
\hfill
\begin{subfigure}{0.41\textwidth}
    \includegraphics[width=\linewidth]{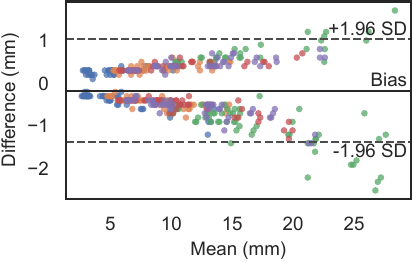}
    \caption{Bland-Altman plot of the difference (system---caliper) vs.~the mean (system and caliper), with 95\% limits of agreement.}
\end{subfigure}
\hfill
\begin{subfigure}{0.13\textwidth}
    \includegraphics[width=\linewidth]{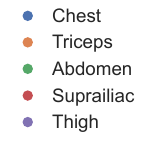}
    \vspace{0.1\textheight}
\end{subfigure}
\caption{Pooled skinfold thickness estimation across all sites and participants.}\label{fig:bodyfat_scatter}
\end{figure}

\paragraph{Overall Accuracy.}
\Cref{tab:eval-main} reports per-site and pooled skinfold thickness error across the five body sites of \Cref{sec:system:sites}.
Pooled across the 15 participants and 5 sites (450 recordings, 4500 paired observations), \system{} attains MAE of 0.54~mm, RMSE of 0.63~mm, \(R^2 = 0.99\), and bias of -0.17~mm in skinfold thickness.
The per-site breakdown exposes which anatomical contexts dominate the residuals rather than absorbing them into a single global metric.
\Cref{fig:bodyfat_scatter} shows the per-observation scatter of \system{} predictions against the caliper ground truth, where each observation inherits the caliper reading of its parent recording, with the Bland-Altman panel quantifying the 95\% limits of agreement.

\begin{table}[t]
\centering
\caption{Per-site and pooled skinfold thickness estimation accuracy. Errors are in millimeters of skinfold thickness.}\label{tab:eval-main}
\begin{tblr}{colspec={lccccccc}}
\toprule
Body site & RMSE (mm) & MAE (mm) & \(R^2\) & Bias (mm) & Diff SD & LoA Lower & LoA Upper \\
\midrule
Chest & 0.40 & 0.36 & 0.99 & -0.08 & 0.39 & -0.85 & 0.69 \\
Triceps & 0.42 & 0.41 & 0.98 & -0.06 & 0.42 & -0.89 & 0.77 \\
Abdomen & 1.03 & 0.93 & 0.98 & -0.44 & 0.94 & -2.28 & 1.41 \\
Suprailiac & 0.51 & 0.48 & 0.99 & -0.14 & 0.50 & -1.11 & 0.83 \\
Thigh & 0.58 & 0.54 & 0.99 & -0.13 & 0.57 & -1.24 & 0.98 \\
\midrule
Pooled & 0.63 & 0.54 & 0.99 & -0.17 & 0.61 & -1.37 & 1.03 \\
\bottomrule
\end{tblr}
\end{table}

\paragraph{Effect of Body Morphology.}
We analyze \system{} accuracy as a function of participant BMI, sex, age, and skinfold thickness range.
The motivation is to verify that the physics-inspired model does not silently degrade for body types underrepresented in the training cohort.

\subsection{Ablation Study}\label{sec:eval:ablation}

We ablate \system{} along three axes: the model architecture, four radar-skin coupling factors (A1--A4), and three geometry and environment factors (A5--A7).
The default deployment posture (\Cref{sec:impl:groundtruth}) places the radar on dry bare skin at a fixed body site, normal to the skin surface, in a quiet indoor environment.

\paragraph{Model Ablation.}
We compare \system{} against two baselines that strip away one half of its design.
The pure-physics baseline keeps the four-medium transfer-matrix forward map and the analytic system-pulse parameterization but disables every neural correction (\(\mathrm{MLP}_{\varepsilon}\), \(\mathrm{MLP}_{\Gamma}\), \(\mathrm{MLP}_{P}\), \(\mathrm{MLP}_{\Gamma^{(0)}}\), \(\mathrm{MLP}_{\mathrm{sys}}\)) and removes the Conv1D encoder.
The pure-NN baseline keeps the same Conv1D encoder and a comparable parameter count but discards the transfer-matrix trunk, mapping each observation directly to \(\{g, d_s, d_f\}\) through a stack of MLPs.
\Cref{tab:eval-ablation-model} reports pooled accuracy across the 15 participants and 5 sites.

\begin{table}[t]
\centering
\caption{Model ablation, pooled across the 15 participants and 5 sites (450 recordings, $ N = 4500 $ paired observations). All values are in millimeters of skinfold thickness, matching standard caliper output.}\label{tab:eval-ablation-model}
\begin{tblr}{colspec={lccccccc}}
\toprule
Model & RMSE (mm) & MAE (mm) & \(R^2\) & Bias (mm) & Diff SD & LoA Lower & LoA Upper \\
\midrule
\system{} (Ours) & 0.63 & 0.54 & 0.99 & -0.17 & 0.61 & -1.37 & 1.03 \\
Pure NN & 2.35 & 1.83 & 0.96 & -1.41 & 1.57 & -4.49 & 1.68 \\
Pure physics & 4.29 & 3.42 & 0.88 & -0.57 & 4.25 & -8.90 & 7.76 \\
\bottomrule
\end{tblr}
\end{table}

\paragraph{Radar-Skin Coupling Ablations (A1--A4).}
This group covers four factors at the radar-skin interface that perturb the contact assumption: standoff distance (A1), clothing (A2), lotion (A3), and skin moisture (A4). Results are summarized in \Cref{tab:eval-ablation-coupling}.

\textit{A1: Standoff Distance.}
Three volunteers are measured at mechanically controlled standoffs of 5, 15, and 30\,mm air gap using metered plastic spacer in addition to the default setting.
MAE rises monotonically with standoff because the growing air-skin reflection competes with the weaker fat/muscle echo (echo~\#3 in the propagation model) and small gap variations rotate the multipath phase by tens of degrees.
At 5\,mm the degradation is moderate ($+0.60$\,mm MAE) because the additional round-trip phase is small relative to the 460\,MHz analysis bandwidth.
At 15\,mm performance degrades substantially ($+1.93$\,mm), and at 30\,mm the air-gap echo overwhelms the tissue response, rendering the estimate unreliable ($+4.38$\,mm).

\textit{A2: Clothing.}
Three volunteers repeat the abdomen measurement with the radar placed over a single dry layer of plain-weave cotton t-shirt fabric (${\sim}2$\,mm, $\varepsilon_r \approx 1.8$).
The fabric reintroduces an air-skin reflection and adds sub-wavelength geometric noise from the weave and seams that violates the planar-layer assumption of the transfer-matrix forward model.
The resulting MAE increase ($+0.81$\,mm) is comparable to the 5\,mm standoff condition, consistent with the interpretation that a loosely draped shirt creates an effective air gap of a few millimeters plus a thin dielectric slab.

\textit{A3: Lotion.}
Three volunteers are measured first with bare skin (baseline) and then with a thin operator-applied layer of Vaseline between the radar face and the skin to simulate the effect of lotion.
The observed MAE change is negligible ($+0.04$\,mm), confirming that a petroleum-jelly-class coupling agent does not perturb the measurement and may even reduce micro-air-gap variability.

\textit{A4: Skin Moisture.}
Three volunteers are measured twice: once at rest with dry skin (baseline) and once lightly perspiring (operator-moistened with physiological saline to a controlled level).
Sweat is a saline solution ($\varepsilon_r \approx 60$--$75$, $\sigma \approx 1.4$\,S/m) that raises the effective permittivity and conductivity of the outermost dermal sub-millimeter.
This shifts the skin Cole-Cole parameters away from the literature values used in the fixed forward model, increasing model mismatch.
The effect is modest ($+0.48$\,mm MAE) because the perturbation is confined to a thin surface film and the petrolatum coupling layer partially isolates the antenna from the sweat.
The physics-loss residual $\lVert h_{\mathrm{obs}} - h_{\mathrm{pred}}\rVert^{2}$ rises by 34\,\% on the post-exercise observations, confirming that the model correctly flags the affected recordings.

\begin{table}[t]
\centering
\caption{Radar-skin coupling ablations on the abdomen site ($N = 15$ participants). Baseline is the default contact, bare-skin setting, marked in gray.}\label{tab:eval-ablation-coupling}
\begin{tblr}{colspec={lllcccc}, row{1} = {font=\bfseries}, row{2,6,8,10} = {bg=gray9}}
\toprule
ID & Ablation & Conditions & RMSE (mm) & MAE (mm) & \(\Delta \)RMSE (mm) & \(\Delta \)MAE (mm) \\
\midrule
\SetCell[r=4]{}A1 & \SetCell[r=4]{}Standoff distance & 0~mm & 0.63 & 0.54 & 0 & 0 \\
 &  & 5~mm & 1.48 & 1.14 & {+0.85} & {+0.60} \\
 &  & 15~mm & 3.17 & 2.47 & {+2.54} & {+1.93} \\
 &  & 30~mm & 6.25 & 4.92 & {+5.62} & {+4.38} \\
\midrule
\SetCell[r=2]{}A2 & \SetCell[r=2]{}Clothing & bare skin & 0.63 & 0.54 & 0 & 0 \\
 &  & dry t-shirt & 1.73 & 1.35 & {+1.10} & {+0.81} \\
\midrule
\SetCell[r=2]{}A3 & \SetCell[r=2]{}Lotion & None & 0.63 & 0.54 & 0 & 0 \\
 &  & Vaseline & 0.67 & 0.58 & {+0.04} & {+0.04} \\
\midrule
\SetCell[r=2]{}A4 & \SetCell[r=2]{}Skin moisture & dry & 0.63 & 0.54 & 0 & 0 \\
 &  & sweat & 1.28 & 1.02 & {+0.65} & {+0.48} \\
\bottomrule
\end{tblr}
\end{table}

\paragraph{Geometry and Environment Ablations (A5--A7).}
This group covers three factors beyond the contact interface that perturb the geometric or environmental assumption: radar angle (A5), surrounding scenario (A6), and pedestrian interference (A7).

\textit{A5: Radar Angle.}
Three volunteers repeats the abdomen measurement with the radar held at four controlled tilt angles ($0^{\circ}$, $15^{\circ}$, $30^{\circ}$, $45^{\circ}$) using a plastic protractor jig that constrains the radar face relative to the skin tangent plane.
At $0^{\circ}$ the planar 1-D transfer-matrix cascade holds exactly.
At $15^{\circ}$ the effective single-pass path length through fat increases by a factor of $1/\!\cos(15^{\circ}) = 1.035$, and the MAE rises only marginally ($+0.27$\,mm) because the Fresnel TE/TM splitting is still negligible at the petrolatum---skin interface.
At $30^{\circ}$ the path-length stretch reaches $1.155\times$ and the TE/TM coefficient divergence becomes appreciable; MAE increases by $+1.01$\,mm.
At $45^{\circ}$ the 1-D assumption breaks down: the path length grows by $1.414\times$, the beam footprint on the skin widens so that the illuminated tissue is no longer laterally homogeneous, and the MAE exceeds 2\,mm ($+1.95$\,mm), crossing the operationally useful threshold.

\textit{A6: Scenario.}
Three volunteers repeat the measurement in three environments: a quiet lab (baseline), a living room with TV and an active Wi-Fi access point at 2\,m, and a residential bedroom with a wooden bed frame and plaster walls.
The bedroom and living-room conditions show no detectable difference from the quiet-lab baseline. Their \(\Delta\)MAE values are within the approximately 0.1 mm within-condition repeatability floor of the baseline abdomen measurement.
This result indicates that, after reference division, the tissue-echo structure in \(\Gamma_{\mathrm{TM\_diff}}(f)\) is effectively decoupled from the surrounding electromagnetic environment under the tested indoor conditions.

\textit{A7: Pedestrian Interference.}
Three volunteers are measured under three motion conditions: a quiet lab (baseline), one person walking a fixed loop at ${\sim}1$\,m from the radar, and one person walking at ${\sim}2$\,m.
The 1 m and 2 m pedestrian conditions show no detectable difference from the quiet-lab baseline, with \(\Delta\)MAE values within the approximately 0.1 mm within-condition repeatability floor of the baseline abdomen measurement. These results indicate that single-pedestrian motion at indoor distances of at least 1 m has no detectable impact on the fat-thickness estimate after windowed temporal denoising.

\begin{table}[t]
\centering
\caption{Geometry and environment ablations on the abdomen site ($N = 15$ participants).
Baseline is the default setting in a quiet lab environment, marked in gray.}\label{tab:eval-ablation-environment}
\begin{tblr}{colspec={X[0.1,l]X[l]X[l]X[c]X[c]X[c]X[c]}, row{1} = {font=\bfseries}, row{2,6,9} = {bg=gray9}}
\toprule
ID & Ablation & Conditions & RMSE (mm) & MAE (mm) & \(\Delta \)RMSE (mm) & \(\Delta \)MAE (mm) \\
\midrule
\SetCell[r=4]{}A5 & \SetCell[r=4]{}Radar angle & $0^\circ$ & 0.63 & 0.54 & 0 & 0 \\
 &  & $15^\circ$ & 1.03 & 0.81 & {+0.40} & {+0.27} \\
 &  & $30^\circ$ & 1.95 & 1.55 & {+1.32} & {+1.01} \\
 &  & $45^\circ$ & 3.17 & 2.49 & {+2.54} & {+1.95} \\
\midrule
\SetCell[r=3]{}A6 & \SetCell[r=3]{}Scenario & lab & 0.63 & 0.54 & 0 & 0 \\
 &  & bedroom & 0.65 & 0.56 & {+0.02} & {+0.02} \\
 &  & living room & 0.66 & 0.58 & {+0.03} & {+0.04} \\
\midrule
\SetCell[r=3]{}A7 & \SetCell[r=3]{}Pedestrian interference & none & 0.63 & 0.54 & 0 & 0 \\
 &  & 1~m & 0.67 & 0.59 & {+0.04} & {+0.05} \\
 &  & 2~m & 0.65 & 0.56 & {+0.02} & {+0.02} \\
\bottomrule
\end{tblr}
\end{table}

\subsection{Body-Fat Calculation}\label{sec:eval:bodyfat}

We evaluate whether \system{} can support downstream whole-body body-fat estimation, rather than only site-specific skinfold prediction.
We use the Jackson---Pollock three-site protocol as the caliper-derived reference because it is the standard workflow that \system{} aims to replace~\cite{jackson1978generalized}.
For each volunteer, the Jackson---Pollock estimate is computed from the caliper-measured doubled-fold skinfolds at the required anatomical sites.
We then compute the corresponding \system{} estimate by replacing each caliper skinfold with the radar-estimated skinfold at the same site.
We directly substitute \system’s caliper-equivalent skinfold estimates into Jackson-Pollock protocol.
This conversion matches the doubled-fold convention used by calipers.

We also compare against a consumer BIA scale to contextualize the practical value of \system{}.
BIA scales estimate body-fat percentage from whole-body electrical impedance and demographic inputs.
They are inexpensive and easy to use, but their estimates are sensitive to hydration, recent meals, skin temperature, and electrode contact~\cite{dehghan2008bia,kyle2004bioelectrical}.
This makes them a convenient consumer baseline, but not a direct replacement for site-specific skinfold assessment.
In contrast, \system{} preserves the anatomical structure of the caliper workflow.
It estimates local subcutaneous thickness at the same sites used by the Jackson---Pollock protocol, but avoids skin pinching and manual fold isolation.

\begin{figure}[t]
    \centering
    \includegraphics[width=0.8\columnwidth]{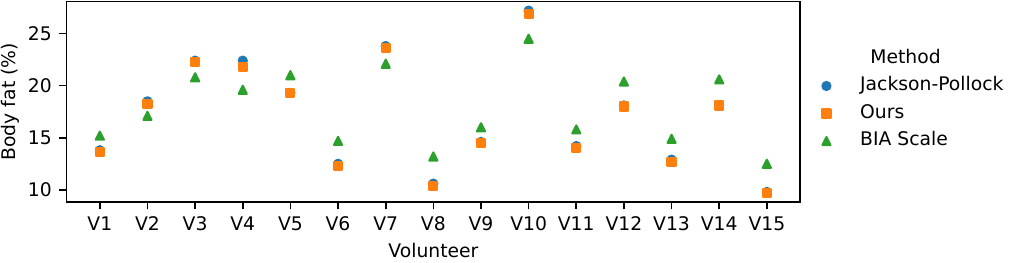}
    \caption{Body-fat percentage estimates for all 15 volunteers. Jackson---Pollock uses caliper-measured three-site skinfolds and serves as the reference. \system{} replaces the caliper measurements with radar-estimated site thicknesses before applying the same protocol. The BIA scale provides a consumer body-composition baseline.}
    \label{fig:bodyfat-comparison}
\end{figure}

\Cref{fig:bodyfat-comparison} compares the three body-fat estimates for all 15 volunteers.
The \system{} estimates closely track the Jackson---Pollock reference across the cohort, while the BIA scale shows larger volunteer-dependent deviations.
Pooled across volunteers, \system{} achieves a mean absolute error of 0.19 percentage points relative to the Jackson---Pollock reference, with RMSE of 0.24 percentage points.
This result shows that sub-millimeter site-level thickness errors translate into small body-fat errors after aggregation across the three skinfold sites.
Small bidirectional residuals at individual sites partially cancel in the three-site sum.

The remaining error is dominated by a mild negative bias.
This bias is consistent with the site-level evaluation in \Cref{tab:eval-main}, where deeper abdominal measurements show slightly larger underestimation.
At these sites, the fat-muscle echo traverses more attenuating tissue, and the fixed Cole---Cole tissue parameters can introduce model mismatch.
Even with this bias, \system{} remains much closer to the Jackson---Pollock reference than the BIA scale in this cohort.
Published inter-rater variability for the Jackson---Pollock protocol ranges from 2.0 to 3.5 percentage points between expert and novice operators~\cite{machado2025skinfold}.
This variability is roughly an order of magnitude larger than the 0.19 percentage-point MAE observed for \system{}.
These results suggest that \system{} can approximate the caliper-based body-fat workflow while removing the need for manual skinfold pinching.

\section{Discussion}
\label{sec:discussion}

\subsection{Practical Implications}
\label{sec:discussion:implications}

\system{} shifts site-specific subcutaneous tissue thickness measurement from a clinic-skill exercise that requires a trained anthropometrist to a self-administered probe that anyone can run on their own body. The site-specific output has direct clinical relevance: subcutaneous fat at the abdomen and suprailiac correlates with metabolic and cardiovascular risk, and longitudinal changes at these sites are a useful marker for tracking response to diet and exercise interventions \cite{jung2016visceral, tchernof2013pathophysiology}. Because \system{} replaces only the caliper, it does not displace whole-body composition tools such as BIA scales or DEXA. The two layers compose: \system{} can supply site-specific thickness inputs to a standalone composition pipeline if such a downstream conversion is desired, and it can do so without operator pinching or current injection.

The UWB hardware class already embedded in consumer smartphones represents a future deployment pathway with no additional hardware cost for end users. A self-administered measurement could be performed by simply holding the phone or watch against the target body site for a few seconds.

\subsection{Privacy and Ethical Considerations}
\label{sec:discussion:privacy}

The same properties that make \system{} attractive, non-intrusive operation, ambient continuous measurement, and reuse of devices the user already carries, also raise privacy and ethical concerns that the system must be designed to address.

\paragraph{Institutional Review Board}
The study protocol was reviewed and approved by the authors' institutional review board under a minimal-risk human-subjects protocol.
All participants provided written informed consent before data collection.
The consent procedure described the body-site measurements, skinfold-caliper measurements, UWB radar recordings, demographic information collected for analysis, and the expected risks of participation.
The primary risk was brief pressure or discomfort during caliper measurements.
Participant data were de-identified before analysis, and all results are reported in aggregate.
Participants were volunteers and did not receive compensation.

\paragraph{Sensitivity of Body Composition Data.}
Subcutaneous fat thickness and its regional distribution are health-relevant attributes with documented potential for discrimination in employment, insurance, and social contexts. Unlike step counts or heart rate, which are now normalized as fitness metrics, body composition carries stigma and can be weaponized against the individual being measured. \system{} treats site-specific thickness readings as protected health information and recommends on-device processing: range profiles and derived parameters should not leave the local device unless the user explicitly opts in to cloud sync or sharing with a healthcare provider.

\paragraph{Implicit and Bystander Measurement.}
Contact-based methods provide a clear consent gesture: the user must step on a scale or hold an electrode. UWB sensing has no such gesture, which is the source of its convenience but also a consent risk. A device performing background measurement may capture the user without an explicit per-session opt-in, and may also pick up household members or visitors who pass within range. We can mitigate this by requiring (1) an explicit one-time opt-in to enable the continuous-measurement mode, (2) on-device person identification so that measurements are only retained for the registered user, and (3) clear, persistent indicators (e.g., a status icon and audible cue) when continuous measurement is active.

\paragraph{Risk for Individuals with Disordered Eating and Body Image Conditions.}
Continuous body composition feedback can be harmful for users with anorexia, bulimia, or body dysmorphic disorder, where access to frequent body metrics is associated with elevated eating disorder symptomatology, including dietary restraint, eating concern, and obsessive use \cite{simpson2017calorie}. \system{} should not be deployed as an unrestricted consumer feature without behavioral safeguards. We recommend (1) age gating, (2) optional cooldown periods between user-visible measurements, (3) the ability to hide individual measurements from the user while still recording trends for clinician review, and (4) integration with existing digital well-being frameworks that flag obsessive checking patterns.

\paragraph{Data Minimization and User Control.}
The signal processing pipeline (Section~\ref{sec:system:pipeline}) yields a compact set of named physical parameters from each measurement; raw range profiles are only needed during model fitting and need not be retained afterwards. We recommend a default configuration that retains only the derived skin and fat thickness parameters per site and discards raw radar data immediately. Users must be able to pause measurement, view all stored records, and delete history at will, in line with established health-data governance principles such as transparency, data minimization, user control, and breach notification, which existing frameworks (HIPAA, GDPR, CCPA) codify but were not originally designed to address the continuous data streams of consumer wearable sensing \cite{doherty2025wearableprivacy}.

\subsection{Limitations}
\label{sec:discussion:limitations}

\paragraph{Penetration Depth.}
UWB signal attenuation in biological tissue sets a practical upper bound on measurable subcutaneous fat thickness. The dominant depth-dependent term is the fat-muscle reflection calculated in \Cref{eq:fm-reflection}.
\begin{equation}
A_{\mathrm{fm}}(d_f) \propto (1-r_{1,2}^{2})\,|r_{2,3}|\, 10^{-\left(2\alpha_s d_s + 2\alpha_f d_f\right)/20},
\label{eq:fm-reflection}
\end{equation}
where \(|r_{2,3}| \approx 0.38\), \(|r_{1,2}| \approx 0.31\), \(\alpha_s = 7.9\)~dB/cm, and \(\alpha_f = 0.98\)~dB/cm at \(f_c = 7.875\)~GHz. Using a nominal skin thickness \(d_s = 2\)~mm and the typical post-averaging SNR of 20~dB, the fixed transmission, reflection, and skin losses consume about \(11.7\)~dB of link margin. The remaining \(8.3\)~dB supports a round-trip fat path of \(8.3/(2\alpha_f) \approx 4.2\)~cm. This gives a first-order detectability limit of \(d_f \approx 40\)~mm for the X7F202 configuration. The bound is above the \(5\)--\(30\)~mm physiological range analyzed in this work, but it suggests that very deep subcutaneous fat may push the fat-muscle reflection toward the commodity receiver noise floor and reduce sensitivity to \(d_f\).

\paragraph{Model Generalizability.} The physics-inspired model is trained and evaluated on a population of 15 participants. Performance on body types significantly outside this distribution (extreme athletes, elderly individuals with muscle wasting, or individuals with edema) is unknown and warrants future study. The cohort BMI range is 18.8–29.1, which does not include the clinical obesity range (BMI \(\leq\) 30). The deployment scenarios discussed in \Cref{sec:discussion:implications}, including visceral risk monitoring and weight-loss intervention tracking, target populations partly outside this validated range; performance at higher BMI is unknown and is a deliberate scope limit of the present feasibility study.

\paragraph{Clothing and Environmental Factors.} Thick winter clothing, metal accessories, and reflective surfaces near the measurement site attenuate or distort UWB signals. Our robustness analysis characterizes these effects up to ${\sim}2$\,mm clothing thickness; highly reflective environments may require compensation.

\paragraph{Radar Bandwidth.} The 460 MHz processing bandwidth is narrow relative to the in-band phase signature of \(d_f\), so identifiability of the individual layer thicknesses \((d_s, d_f)\) from the radar alone is weak. The system therefore reports the doubled-fold sum \(2(d_s+d_f)\) jointly with caliper supervision, not separately recovered layer parameters. Disentangling layers would require either wider bandwidth (e.g., the full 750 MHz TX BW or multi-band fusion) or per-subject ultrasound priors on \(d_s\).

\subsection{Future Work}
\label{sec:discussion:future}

\paragraph{Smartphone-Native Deployment.}
Our prototype uses a stand-alone commodity UWB module to control hardware parameters precisely during evaluation. The natural next step is to port the signal processing and inference pipeline onto a phone-grade UWB chipset and characterize accuracy under the constraints of phone-grade transmit power, antenna placement, and intermittent radio access. This step would convert \system{} from a dedicated device into a software feature on hardware users already own.

\paragraph{Personalization with Minimal Calibration.}
Body habitus, posture, and tissue composition vary enough across individuals that a single population-level model leaves accuracy on the table. A short per-user calibration, for example, a one-time multi-site skinfold reference, could anchor the regression model to that user's baseline and substantially reduce error on subsequent contactless measurements. We will study calibration sample-efficiency: how few reference points are needed before the personalized model meaningfully outperforms the population model.

\paragraph{Multi-Site Fusion for Fat-Distribution Maps.}
The current evaluation treats each body site independently. Combining measurements across multiple sites within a single session could yield a regional fat-distribution map and expose site-to-site correlations that a single-site reading hides. This direction also opens up tracking fat redistribution over time, which is clinically meaningful for interventions targeting visceral versus subcutaneous fat.

\paragraph{Longitudinal Health-Outcome Studies.}
The strongest test of \system{}'s clinical value is whether continuous, longitudinal measurements detect health-relevant change earlier or more reliably than current standards. We plan a multi-month deployment in cohorts undergoing structured interventions (diet, resistance training, weight-loss medication) to assess whether \system{} can quantify intervention response with finer temporal resolution than weekly or monthly clinic visits.

\section{Conclusion}
\label{sec:conclusion}

We presented \system, the first system to use commodity UWB radar as a non-intrusive replacement for the skinfold caliper. By leveraging the dielectric contrast among skin, fat, and muscle, and by fitting a physics-inspired model directly to the complex baseband CIR of a Novelda X7F202 commercial UWB module, \system{} recovers site-specific caliper-equivalent skinfold thickness without skin pinching, without current injection, without a second operator, and without dedicated clinical hardware. Our evaluation across 15 participants against skinfold caliper ground truth demonstrates a pooled mean absolute error of $0.54$\,mm in skinfold thickness across $15$ participants and $5$ anatomical sites. The growing prevalence of UWB in consumer smartphones, smartwatches, and item trackers positions \system{} as a practical path toward ambient, longitudinal body composition monitoring accessible to everyday users. As UWB hardware continues to proliferate in the consumer ecosystem, \system{} establishes the sensing feasibility and system design principles necessary to transform this technology into a ubiquitous health monitoring tool.

\bibliography{references}

\end{document}